\documentclass[aps, prx, reprint,footinbib,floatfix,superscriptaddress]{revtex4-2}
\usepackage{lipsum}
\usepackage{graphicx}
\usepackage{indentfirst}
\usepackage{braket}
\usepackage{float}
\usepackage{amsmath}
\usepackage{physics}
\usepackage{amssymb}
\usepackage{CJK}
\usepackage{esint}
\usepackage{color}
\usepackage[T1]{fontenc}
\usepackage{amsfonts}
\usepackage{footmisc}
\usepackage{scrextend}
\usepackage{multirow}
\usepackage[outdir=./]{epstopdf}
\usepackage{xcolor}
\usepackage[english]{babel}
\usepackage{url}
\usepackage{comment}
\usepackage{array}
\usepackage{subfiles}
\usepackage{tikz}
\usetikzlibrary{shapes,arrows}
\usepackage{mathrsfs}
\usepackage{mathtools}
\usepackage[mathscr]{eucal}
\usepackage{multirow}
 \usepackage{booktabs} 

\usepackage[hyperfootnotes=false]{hyperref}
\usepackage{cleveref}
\definecolor{darkblue}{rgb}{0,0,0.5}
\hypersetup{
colorlinks=true,
linkcolor=black,
filecolor=blue,
citecolor=darkblue,  
urlcolor=cyan,
}

\usepackage{trimclip}

\makeatletter
\DeclareRobustCommand{\shortto}{%
  \mathrel{\mathpalette\short@to\relax}%
}

\newcommand{\short@to}[2]{%
  \mkern2mu
  \clipbox{{.5\width} 0 0 0}{$\m@th#1\vphantom{+}{\shortrightarrow}$}%
  }
\makeatother

\usepackage{pict2e,picture,graphicx}

\makeatletter
\DeclareRobustCommand{\Arrow}[1][]{%
\check@mathfonts
\if\relax\detokenize{#1}\relax
\settowidth{\dimen@}{$\m@th\rightarrow$}%
\else
\setlength{\dimen@}{#1}%
\fi
\sbox\z@{\usefont{U}{lasy}{m}{n}\symbol{41}}%
\begin{picture}(\dimen@,\ht\z@)
\roundcap
\put(\dimexpr\dimen@-.7\wd\z@,0){\usebox\z@}
\put(0,\fontdimen22\textfont2){\line(1,0){\dimen@}}
\end{picture}%
}
\makeatother

\def\be{\begin{equation}}
\def\ee{\end{equation}}
\def\ba{\begin{eqnarray}}
\def\ea{\end{eqnarray}}
\def\bal{\begin{equation}\begin{aligned}}
\def\eal{\end{aligned}\end{equation}}

\def\bp{\begin{pmatrix}}
\def\ep{\end{pmatrix}}

\usepackage{bm}

\newcommand{\calA}{{\cal A}}

\newcommand{\calB}{{\cal B}}
\newcommand{\calC}{{\cal C}}

\newcommand{\calE}{{\cal E}}

\newcommand{\calF}{{\cal F}}

\newcommand{\calL}{{\cal L}}

\newcommand{\calV}{{\cal V}}

\newcommand{\calX}{{\cal X}}

\newcommand{\1}{^{(1)}}

\newcommand{\QZ}[1]{{{\textcolor{black}{#1}}}}

\newcommand{\QZZ}[1]{{{\textcolor{black}{#1}}}}

\newcommand{\hw}[1]{{{\textcolor{black}{#1}}}}
\newcommand{\hww}[1]{{{\textcolor{black}{#1}}}}

\usepackage[normalem]{ulem}

\begin{document}
\title{Theory of quantum comb enhanced interferometry}
\begin{abstract}
Optical frequency combs, named for their comb-like peaks in the spectrum, are essential for various sensing applications. As the technology develops, its performance has reached the standard quantum limit dictated by the quantum fluctuations of coherent light field. Quantum combs, with their quantum fluctuation engineered via squeezing and entanglement, are the necessary ingredient for overcoming such limits. We develop the theory for designing and analyzing quantum combs, focusing on dual-comb interferometric measurement. Our analyses cover both squeezed and entangled quantum combs with division receivers and heterodyne receivers, leading to four protocols with quantum advantages scalable with squeezing/entanglement strength.
\QZ{In the spectroscopy of a single absorption line, the division receiver with the squeezed comb suffers from \QZZ{entanglement-mismatching-induced amplified noise}, while the other three protocols demonstrate a surprising robustness to loss at a few comb lines. Such a unique loss-robustness of a scalable quantum advantage has not been found in any traditional quantum sensing protocols.}

\end{abstract}
\author{Haowei Shi}
\email{haow.shi@gmail.com}
\address{
Ming Hsieh Department of Electrical and Computer Engineering, University of Southern California, Los
Angeles, California 90089, USA
}

\author{Quntao Zhuang}
\email{qzhuang@usc.edu}

\address{
Ming Hsieh Department of Electrical and Computer Engineering, University of Southern California, Los
Angeles, California 90089, USA
}
\address{
Department of Physics and Astronomy, University of Southern California, Los
Angeles, California 90089, USA
}
\maketitle

Frequency combs~\cite{hansch2006,hall2006} refer to the state of light with a spectrum consisting of a comb of equally-spaced discrete lines. Dual-comb interferometry, which interferes two frequency combs with slightly different comb line frequency spacings, has emerged to provide the state-of-the-art performance in \QZ{various applications}~\cite{Picque2019, Coddington:16, Fortier2019,Martin-Mateos_imagingOptica, Vicentini2021,Coddington2009RapidAP, kippenberg_ranging2011, Lukashchuk2022, Muh_ranging2018, Caldwell2022}. \QZ{Continuous efforts have pushed dual-comb interferometry systems} close to the \QZ{quantum shot noise limit.}
To further enhance \QZ{its} signal-to-noise ratio (SNR), \hw{there have been extensive efforts, e.g. frequency averaging~\cite{walsh2023unlocking} and frequency multiplexing~\cite{newbury2010sensitivity}, whereas} quantum engineering of the combs is necessary \hw{to break the fundamental limit}. Ref.~\cite{shi2023entanglement} proposes the \QZ{intra-comb-line} two-mode squeezing \QZ{centered slightly off from} each comb line for heterodyne detection. 
Ref.~\cite{herman2024squeezed} avoids the need of the frequency offset with a division receiver, and experimentally demonstrates squeezing \QZ{effects} in the weak local oscillator (LO) limit. Ref.~\cite{hariri2025entangled} designs a cross-comb-line entanglement structure and experimentally demonstrates quantum advantages for heterodyne detection \QZ{under a low sample power constraint}. Despite the progress, a unified theory \QZ{model to} benchmark quantum advantage is missing.

\begin{figure}[b]
    \centering
    \includegraphics[width=\linewidth]{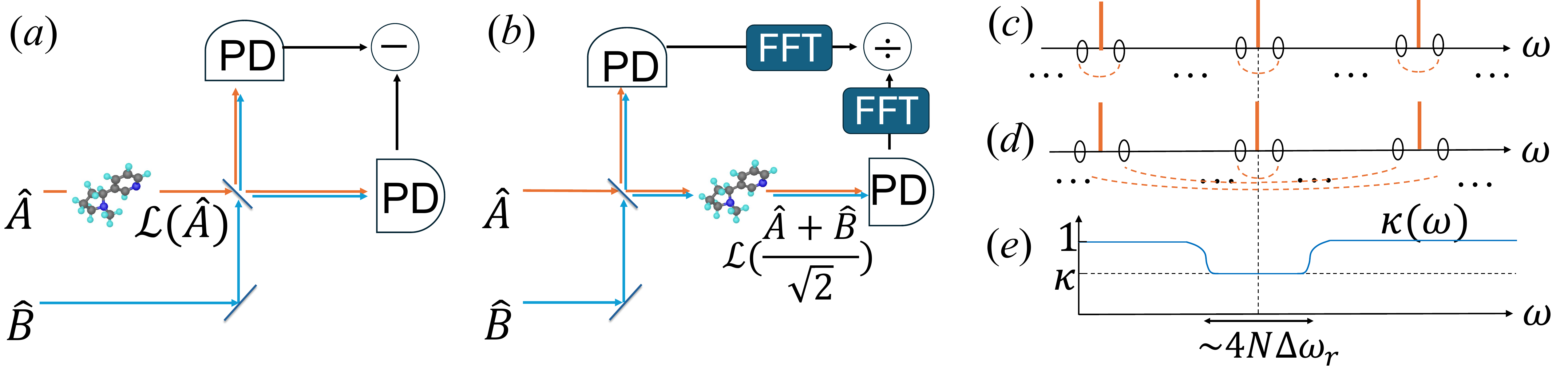}
    \caption{{ Schematic of dual-comb spectroscopy.} (a) Heterodyne-receiver-based DCS. 
    (b) Division-receiver-based DCS. 
    We consider two classes of quantum comb engineering: (c) Intra-comb-line squeezing: squeezed pairs are centered around each line; (d) Cross-comb-line entanglement: squeezed pairs are centered around the carrier frequency. (e) Schematic of a spectrum with a single absorption line. \QZ{We assume the absorption window covers the whole relevant noise sideband around the absorption line.} 
    \label{fig:schematic}
    }
\end{figure}

In this work, we develop the unified theory for quantum frequency combs and its SNR in dual-comb interferometry, focusing on the application of dual-comb spectroscopy (DCS). We consider quantum combs with strong `classical' comb lines, different from Refs.~\cite{pysher2011parallel,yang2021squeezed,shen2025highly,wang2025large,wilken2024broadband,myilswamy2023time,dalvit2024quantum} \QZZ{where the combs are weak}, to enable quantum metrology advantage over the classical counterparts in DCS.
We analyze four protocols---combining intra-comb-line squeezing~\cite{shi2023entanglement,herman2024squeezed} or cross-comb-line entanglement~\cite{hariri2025entangled} \QZZ{with} heterodyne or division receivers, including three existing ones~\cite{shi2023entanglement,herman2024squeezed,hariri2025entangled} and an additional entangled-enhanced division receiver scheme (see \QZ{Fig.~\ref{fig:schematic}} and Table~\ref{table_summary}). \QZ{Beyond-constant quantum advantage generally requires both combs quantum engineered, with the exception of heterodyne detection under sample power constraint, where a single quantum comb is sufficient.} We show that the division receiver with the squeezed comb as in Ref.~\cite{herman2024squeezed} suffers from \QZZ{entanglement-mismatching-induced amplified noise}, even when detecting a single absorption line among transparent backgrounds; while the other three protocols demonstrate a surprising robustness to loss---arbitrary amount of loss in a small portion of the frequency lines does not decrease the SNR advantage. Such a loss-robustness of a scalable quantum advantage has not been found in any quantum sensing scenarios, cf. the squeezed-based interferometer vulnerable to loss~\cite{demkowicz2013fundamental,frascella2021overcoming}.

\section{Dual comb interferometry}
\begin{figure}[t]
    \centering
    \includegraphics[width=\linewidth]{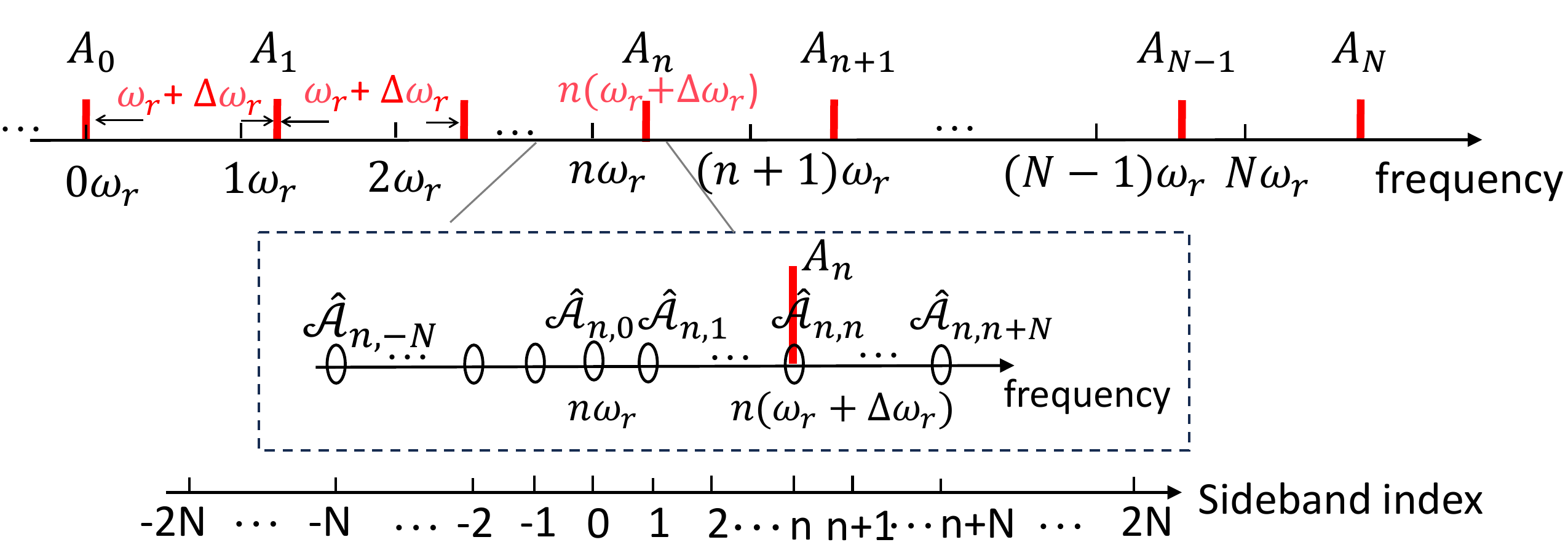}
    \caption{Schematic of the mode definitions. The comb lines are denoted by $A_n$, located at frequency $n(\omega_r+\Delta \omega_r)$, for $-N\le n\le N$. For the noise modes $\{\hat \calA_{n,m}\}$, the first subscript $n$ indexes the comb line, while the second subscript $m$ indexes the detuning of the sideband noise mode from $n\omega_r$, \hww{which gives frequency $n\omega_r+m\Delta\omega_r$.}\QZZ{The modes for comb $B$ are defined in the same fashion, with the difference that the comb lines $B_n$ are at the $\hat\calB_{n,0}$ mode with frequency $n\omega_r$.}}
    \label{fig:schematic_mode_main}
\end{figure}

A quantum-engineered frequency comb can be described by a field operator $\hat A$, with a classical mean value $\expval{\hat A}$ and quantum fluctuations $\QZ{\Delta\hat{A}\equiv}\hat A-\expval{\hat A}$ \QZ{(see Appendix A)}. In a \QZ{dual comb interferometry set-up}, two combs are involved with different frequency spacings \QZ{$\omega_A=\omega_r+\Delta \omega_r$ and $\omega_B=\omega_r$}---in terms of the mean values
\bal 
\expval{\hww{\hat C}(t)}&=\frac{1}{\sqrt{T}}\sum_{n=-N}^N e^{-i n \omega_r t} \hww{C_n},
\label{eq_combAB_mean_main_text}
\eal 
where $\hww{C}=A, B$ representing two combs and the summation of the comb-line index $n$ is over a total number of $M=2N+1$ comb lines, each with amplitude $A_n$ or $B_n$. \QZ{The interference between the two combs, in the form of $\hat{A}^\dagger \hat B$, will generate both high-frequency signals at $(n-n^\prime)\times \omega_r$ between comb lines with different indices $n,n^\prime$, and low-frequency signals at $n\Delta \omega_r$ between each pair of $A_n$ and $B_n$. }
In practice, detectors will capture the low-frequency signals at frequencies $\{n\Delta \omega_r\}_{n=-\QZ{N}}^\QZ{N}$ to learn the absorption across the entire frequency band $\{n\omega_r\}_{n=-\QZ{N}}^\QZ{N}$. \QZ{Therefore, we decompose the quantum fluctuations into a discrete set of modes 
\bal 
\Delta \hww{\hat C}(t)&=\frac{1}{\sqrt{T}}\sum_{n=-N}^N\sum_{m=-2N}^{2N} e^{-i (n\omega_r+m\Delta\omega_r) t}\hat \calC_{n,m},
\label{eq_noise_main}
\eal 
for both $\hww{C}=A, B$, \hww{and $\cal C=\calA, \calB$}. \QZ{The modes $\hat \calA_{n,m}$ and $\hat \calB_{n,m}$ will be quantum engineered. A full schematic of the modes can be found in Fig.~\ref{fig:schematic_mode_main}.}
}


In a linear absorption spectroscopy measurement, a sample can be modeled by frequency-dependent transmissivity $\kappa(\omega)$ and phase $\theta(\omega)$, \QZ{where each frequency mode goes through a bosonic loss channel $\calL$~\cite{weedbrook2012gaussian}: $\hat A(\omega)\to \sqrt{\kappa(\omega)}e^{i\theta(\omega) } \hat A(\omega) +\sqrt{1-\kappa(\omega)} \hat v(\omega)$.}
Although the analyses in \cite{supp} includes thermal noise, we will ignore it throughout the main text and thus $\hat v(\omega)$ denotes vacuum. 
\QZ{For simplicity, we consider the channel to be uniform within the neighborhood of each comb, $\kappa(n\omega_r+m \Delta \omega_r)\equiv \kappa_n$ and $\theta(n\omega_r+m \Delta \omega_r)\equiv \theta_n$, such that the mean and noise
\be 
\hww{C_n^\prime}= \sqrt{\kappa_n}e^{i\theta_n}\QZZ{C_n}, \, \hww{\hat \calC^\prime_{n,m}}= \sqrt{\kappa_n}e^{i\theta_n} \hww{\hat \calC_{n,m}}+{\cdots},
\label{loss_channel}
\ee 
with `$\cdots$' denoting the loss-induced terms.
}

To obtain information about the sample, we consider two DCS set-ups. 
In the first case (see Fig.~\ref{fig:schematic}a), we pass one of the comb (e.g. $\hat{A}$) through the sample and then combine the two combs for photo-detection. The difference of the photon-current on the two detectors is used to extract the spectrum of absorption, forming a heterodyne detection~\cite{shi2023entanglement,hariri2025entangled}.
The quantum operator describing the heterodyne in the spectral domain is
\be 
\hat{d}_m\equiv \int  e^{im \Delta\omega_r t}\left\{\calL[\hat A(t)]^\dagger \hat B(t)+\hat B^\dagger(t)\calL[\hat A(t)]\right\} \hww{dt}.
\label{difference_current_main}
\ee 
Keeping the low frequency terms and combining Eqs.~\eqref{eq_combAB_mean_main_text}-\eqref{loss_channel}, the mean and leading-order noise
\begin{align} 
&\expval{\hat{d}_m}=\sqrt{\kappa_m}e^{i\theta_n} A_m B_m^\star+\sqrt{\kappa_{-m}}e^{-i\theta_n} A_{-m}^\star B_{-m}.
\label{d_m_het}
\\
&
\Delta \hat{d}_m=
\sum_{n=-N}^N A_n^{\prime \star}\hat \calB_{n,n+m}+A_n^\prime\hat \calB_{n,n-m}^{\dagger}
 \nonumber\\
 &\qquad \qquad \qquad +B_n^\star\hat \calA^\prime_{n,m}+B_n\hat \calA_{n,-m}^{\prime\dagger}.
 \label{noise_maintext}
\end{align}
Eq.~\eqref{d_m_het} indicates that heterodyne DCS is capable of detecting both the phase and absorption induced by the sample;
While Eq.~\eqref{noise_maintext} shows its sensitivity to phase fluctuation, 
\QZZ{as it involves $A^\prime_n$ and $\hat \calA^\prime_{n,\pm m}$ which depend on phase $\theta_n$ via Eq.~\eqref{loss_channel}}.

In the second case (see Fig.~\ref{fig:schematic}b), we combine the two combs $\hat A(t)$ and $\hat B(t)$ \QZ{to produce $\hat A'(t)=\frac{\hat A(t)+\hat B(t)}{\sqrt{2}}$ and $
\hat B'(t)=\frac{\hat A(t)-\hat B(t)}{\sqrt{2}}\,$} first, and then pass one of the combined beam through the sample, before the final photo-detection, \QZ{$\hat I_A(t)
=\calL[\hat A^{\prime}(t)]^\dagger\calL[\hat A'(t)]$ and $\hat I_B(t)=\hat B^{\prime\dagger}(t)\hat B'(t)$}. In this scenario, as one of the arm (the upper arm in Fig.~\ref{fig:schematic}b) serves as a reference without knowledge about the sample, it is natural to adopt a division receiver scheme---taking the ratio of photo current spectra, \QZZ{$\hat{r}_m\equiv-\hat I_A(m\Delta \omega_r)/\hat I_B(m\Delta \omega_r)$ with} $\hat I_{\QZZ{C}}(m\Delta \omega_r)\equiv \int e^{im \Delta\omega_r t} \hat I_{\QZZ{C}}(t) dt$ \QZZ{for $C=A,B$,} to estimate sample absorption~\cite{herman2024squeezed}. \QZ{Indeed, their mean values are
\begin{subequations}
\ba 
&\!\!\!\!\!\!\expval{\hat I_A(m\Delta \omega_r)}=&\frac{1}{2}\left(\kappa_m A_m B_m^\star+\kappa_{-m} A_{-m}^\star B_{-m}\right),
\\
&\!\!\!\!\!\!\expval{\hat I_B(m\Delta \omega_r)}=&\frac{1}{2}\left(-A_mB_m^\star-A_{-m}^\star B_{-m}\right).
\ea 
\label{I_mf_mean_main}
\end{subequations}
\QZZ{Therefore, the mean of ratio $\expval{\hat{r}_m}\simeq -\kappa_m$ provides a direct estimation of the absorption, if the amplitudes $A_{m}=B_{m}=0$ for all $m<0$.} \QZZ{The absence of $\theta_n$ in}
Eqs.~\eqref{I_mf_mean_main} indicates that division receiver DCS cannot measure sample-induced phase due to the self-interference.} 
At the same time, it is robust to phase fluctuation. The noise of a division receiver \QZZ{comes} from both current spectra:
\begin{align}
&\Delta \hat I_A(m \Delta \omega_r)={\cdots}+
\nonumber
\\
&\frac{1}{2}\sum_{n=-N}^N \!\kappa_n \! \left[A_n^\star (\hat \calA_{n,n+m}+\hat \calB_{n,n+m}) \!+\! B_n^\star(\hat \calA_{n,m}+\hat \calB_{n,m}) \right]\nonumber
\\
&+\frac{1}{2}\!\sum_{n=-N}^N \!\kappa_n \!\left[\! A_n (\hat \calA_{n,n-m}^\dagger\!+\!\hat \calB_{n,n-m}^\dagger)\! +\! B_n(\hat \calA_{n,-m}^\dagger\!+\!\hat \calB_{n,-m}^\dagger) \right],
\label{eq:photocurrent_spectrum_A_main}
\end{align}
\begin{align}
&\Delta \hat I_B(m \Delta \omega_r)=
\nonumber
\\
&\frac{1}{2}\sum_{n=-N}^N \left[A_n^\star (\hat \calA_{n,n+m}-\hat \calB_{n,n+m})-B_n^\star(\hat \calA_{n,m}-\hat \calB_{n,m}) \right]
\nonumber
\\
&+\left[A_n (\hat \calA_{n,n-m}^\dagger-\hat \calB_{n,n-m}^\dagger)-B_n(\hat \calA_{n,-m}^\dagger-\hat \calB_{n,-m}^\dagger) \right]\,.
\label{eq:photocurrent_spectrum_B_main}
\end{align}
where $\cdots$ represents loss-induced terms.
\QZZ{
The fluctuation of the ratio $\Delta \hat{r}_m\propto \Delta\hat I_A(m\Delta \omega_r)+\expval{\hat r_m} \Delta\hat I_B(m\Delta \omega_r)$ combines the above,
\begin{align}
&\Delta \hat{r}_m
\propto \sum_{n=-N}^N (\kappa_n+\expval{\hat r_m})
\left[
A_n^\star \hat \calA_{n,n+m}+
A_n \hat \calA_{n,n-m}^\dagger
\right.
\nonumber
\\
&
\left.  \quad \quad \quad \quad \quad \quad \quad \quad \quad \quad \quad \quad 
+B_n^\star\hat \calB_{n,m}
+B_n\hat \calB_{n,-m}^\dagger
\right]
\nonumber
\\
&
+\frac{1}{2}\sum_{n=-N}^N 
(\kappa_n-\expval{\hat r_m})
\left[
A_n^\star \hat \calB_{n,n+m}
+A_n \hat \calB_{n,n-m}^\dagger
\right.
\nonumber
\\
&
\left.\quad \quad \quad \quad \quad \quad \quad  \quad \quad \quad 
+B_n^\star\hat \calA_{n,m} 
+B_n\hat \calA_{n,-m}^\dagger
\right]+\cdots,
\label{IA_IB_division_sum_main}
\end{align}
where we have ignored some vacuum related terms.
}

\section{Quantum engineering of frequency combs}
In both scenarios, we consider the quantum engineering of the combs to suppress the quantum shot noise and enhance the SNR. While the above framework paves the way for the analyses of general quantum engineering, we focus on two-mode squeezing in recent experiments~\cite{herman2024squeezed,hariri2025entangled}. We use the wording ‘intra-comb-line squeezing’ and
 ‘cross-comb-line entanglement’ to specify the pairing structure of the
 noise modes.

The first type of quantum comb involves intra-comb-line squeezing~\cite{shi2023entanglement,herman2024squeezed}. As shown in Fig.~\ref{fig:schematic}(c), two-mode squeezing is applied on sideband frequency pairs centered around each comb line individually, at multiples of $\Delta \omega_r$ apart.  
\QZ{While it is well-known that single-mode squeezing cannot enhance heterodyne detection beyond $3$dB, two-mode squeezing can be utilized to overcome such limits~\cite{collett1987quantum}.} \QZZ{Such a squeezing structure can be generated via adopting a classical comb as the pump in a nonlinear fiber~\cite{herman2024squeezed} or other systems~\cite{pinel2011generation,medeiros2014full}, with the remnant of the pump comb lines naturally serving as the nonzero amplitudes in the quantum comb.}

\begin{figure}[t]
    \centering
    \includegraphics[width=\linewidth]{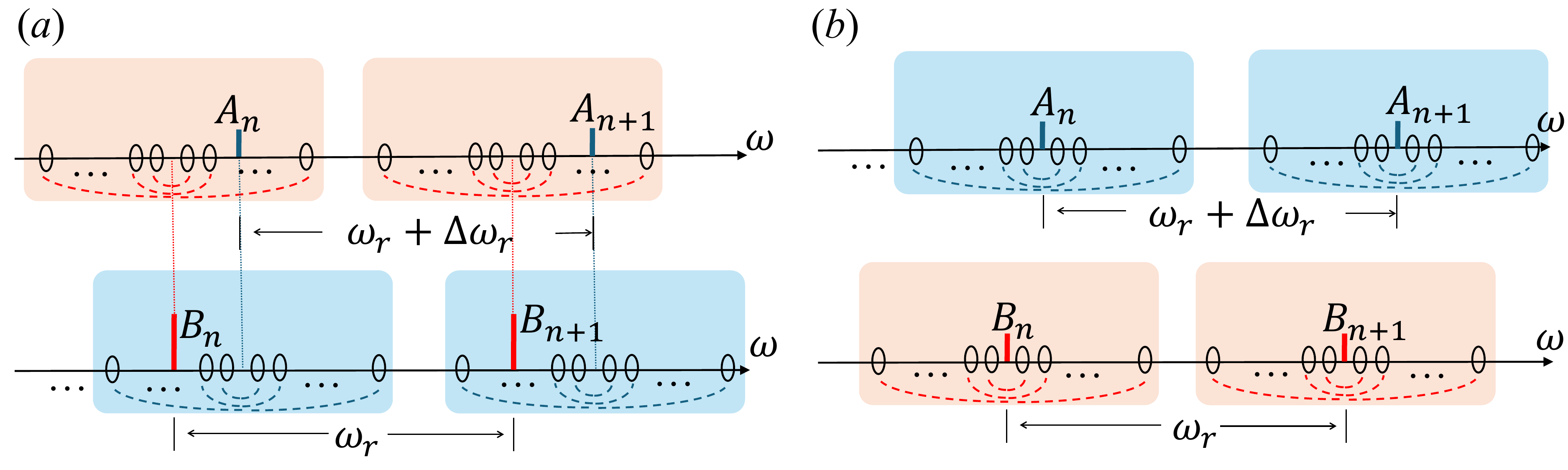}
    \caption{
    Two cases of intra-comb-line squeezing. (a) cross-referenced squeezing pairs, where comb $A$ noises are paired around lines in comb $B$ and vice versa; (b) self-referenced squeezing pairs, where comb $A$ noises are paired around lines in comb $A$ itself and similar for comb $B$.
    \label{fig:sqz_structure}
    }
\end{figure}

To suppress the fluctuation \QZ{in heterodyne detection}, Eq.~\eqref{noise_maintext} indicates a two-mode squeezing \QZ{(see Appendix B)} between each cross-referenced pair of modes $\hat\calA_{n,\pm m}$ (similarly $\hat \calB_{n,n\pm m}$)~\cite{shi2023entanglement}, leading to the cross-referenced intra-comb-line squeezing structure shown in Fig.~\ref{fig:sqz_structure}a. \QZ{Due to the pairwise summation over all $M$ comb lines in Eq.~\eqref{noise_maintext}, when a small portion of the absorption $\kappa_n<1$, the vacuum noise remains small. These properties makes intra-comb-line squeezing enhanced heterodyne detection robust to partial loss among the spectrum.}

To suppress the fluctuation in division detection, ideally one hopes to suppress the fluctuation in both $\Delta \hat I_A(m \Delta \omega_r)$ and $\Delta \hat I_B(m \Delta \omega_r)$ in Eqs.~\eqref{eq:photocurrent_spectrum_A_main} and \eqref{eq:photocurrent_spectrum_B_main} individually; however, the direct mode pairing in $\Delta \hat I_A(m \Delta \omega_r)$ involves both \QZZ{the pair} $\hat \calA_{n,n\pm m}$ and \QZZ{the pair} $\hat \calA_{n,\pm m}$ for each $n,m$, requiring incompatible two-mode squeezing set-ups for simultaneously suppression (similarly, $\hat \calB_{n,n\pm m}$ and $\hat \calB_{n,\pm m}$). Thus, one would only hope to suppress the fluctuation of the ratio $\hat{r}_m$ \QZZ{in Eq.~\eqref{IA_IB_division_sum_main}}.
Indeed, if \QZZ{$\kappa_n=\overline\kappa$ (thus $\expval{\hat r_m}=\overline\kappa$) is uniform} for all comb lines $n$, \QZZ{the second summation in Eq.~\eqref{IA_IB_division_sum_main} is canceled due to the pre-factor $(\kappa_n-\expval{\hat r_m})=0$}, leaving only \hww{the self-referenced noise terms} $\hat \calA_{n,n\pm m}$ and $\hat \calB_{n,\pm m}$ \QZZ{in the first summation of Eq.~\eqref{IA_IB_division_sum_main}} needed to be two-mode squeezed for noise suppression. Therefore, the side-band two-mode squeezing centers at frequencies $n (\omega_r+\Delta\omega_r)$, aligned with the comb lines $A_n$~\cite{herman2024squeezed}, as shown in Fig.~\ref{fig:sqz_structure}b. However, the cancellation does not generally happen, unless for uniform absorption across all frequency. A single heterogeneous absorption line \QZZ{$\kappa_\ell\neq \overline\kappa$} \QZZ{introduces substantial noise for the corresponding estimator $\hat{r}_\ell$, as $\expval{\hat r_\ell}\neq \kappa_n$ for all $n\neq \ell$. Consequently, all the noise terms with $n\neq \ell$ in the second summation in Eq.~\eqref{IA_IB_division_sum_main} remain nonzero as $(\kappa_{n\neq \ell}-\expval{\hat r_\ell})\neq 0$. If one entangles the self-referenced pairs $\{(\hat \calA_{n,n + \ell}, \hat \calA_{n,n - \ell})\}_{n}$ according to Fig.~\ref{fig:sqz_structure}b, the cross-referenced terms $B_n^\star \hat \calA_{n,\ell} 
+B_n\hat \calA_{n,-\ell}^\dagger$ in the second summation of Eq.~\eqref{IA_IB_division_sum_main} simply sums two independent modes in thermal states with amplified noises, and vice versa. Similar imperfect cancellation of noises happens for $\hat \calB_{n,\pm \ell}$. We define such amplified noises as entanglement-mismatching-induced noise.} This represents a loss-sensitive intra-comb-line squeezing enhanced division detection.

The second type of quantum comb adopts cross-comb-line entanglement~\cite{hariri2025entangled}. 
\QZ{For heterodyne detection noise in Eq.~\eqref{noise_maintext}, one can further pair the positive and negative frequencies as
\small
\begin{align} 
&\Delta \hat{d}_m=\sum_{n=1}^N(B_n^\star\hat \calA^\prime_{n,m}+B_{-n}\hat \calA_{-n,-m}^{\prime\dagger}+B_n\hat \calA_{n,-m}^{\prime\dagger}+B_{-n}^\star\hat \calA^\prime_{-n,m})
\nonumber
\\
&+(A_n^{\prime \star}\hat \calB_{n,n+m}+A_{-n}^\prime\hat \calB_{-n,-n-m}^{\dagger}+A_n^\prime\hat \calB_{n,n-m}^{\dagger}+A_{-n}^{\prime \star}\hat \calB_{-n,-n+m}),
 \label{noise_maintext_pairing}
\end{align}
\normalsize
where we ignored the $n=0$ term in the summation. Here, the pairing of each positive frequency mode $\hat \calA_{n,m}$ and its negative frequency mode $\hat \calA_{-n,-m}$ in a TMSV (similarly $\hat \calB_{n,n+m},\hat \calB_{-n,-n-m}$)~\cite{hariri2025entangled} suppresses the fluctuation (under matching phase), leading to the cross-comb-line entangled comb in Fig.~\ref{fig:schematic}d centered around the carrier frequency (the center line), and across the entire frequency domain of multiples of $\omega_r$.
Similarly, in the division detection approach, both $\Delta \hat I_A(m \Delta \omega_r)$ and $\Delta \hat I_B(m \Delta \omega_r)$ (see Eqs.~\eqref{eq:photocurrent_spectrum_A_main} and \eqref{eq:photocurrent_spectrum_B_main}) can be individually suppressed by the cross-comb-line entangled comb, \QZZ{therefore avoiding the entanglement-mismatching-induced noise in the intra-comb-line case}. However, as the two-mode squeezed pairs, $\hat \calA_{n,m}$ and $\hat \calA_{-n,-m}$, go through generally different transmissivities, $\kappa_n\neq \kappa_{-n}$, 
the asymmetry introduces amplified noise \QZZ{for both division receiver and heterodyne}. Nevertheless, such \hww{asymmetry-induced} amplified noise is local around only the absorption line and its asymmetric image, independent on $M$, thus much more robust than the division receiver with intra-comb-line squeezing case.
}

Note that quantum engineering can be applied to both combs in DCS, and in fact often required to guarantee quantum advantage (see Table~\ref{table_summary}). Moreover, the two-mode squeezing may be frequency-dependent, as we analyze in \cite{supp}.
For simplicity, in the main text, we consider uniform comb lines, $A_n=A$ and $B_n=B$, and uniform two-mode squeezing gain $G_A\ge1$ and $G_B\ge1$ correspondingly. Here the squeezing gain $G$ describes the suppression of Einstein–Podolsky–Rosen (EPR) quadrature variances $1/G$ below the vacuum limit (see Appendix B). The quantum combs degenerate back to classical ones when the squeezing gains $G=1$.

\section{Performance}
To illustrate and compare the quantum enhancement, we consider probing a simple sample with a single absorption line at \QZ{unknown} frequency $m \omega_r$ (see Fig.~\ref{fig:schematic}e). Mathematically, this means $\kappa_m=\kappa<1$ and phase shift $\theta_m=\theta\neq 0$ among transparent backgrounds $\kappa_{n}=1$ and $\theta_{n}=0$ for all $n\neq m$. \QZ{Here $\kappa_m$ and $\theta_m$ apply to the frequency region $[m \omega_r-2N\Delta \omega_r,m \omega_r+2N\Delta \omega_r]$, affecting both the comb line and sideband.} We benchmark the local SNR using the Fisher information for parameter $\sqrt{\kappa}$. Below, we summarize our results in the single-absorption-line example, while the details of the analyses, \QZZ{global SNR analyses, phase sensing}, and the general case of arbitrary sample absorption and non-uniform combs can be found in \cite{supp}. We will focus on the $M\gg G_A, G_B$ limit to simplify the expressions.

\begin{figure}[t]
    \centering
    \includegraphics[width=\linewidth]{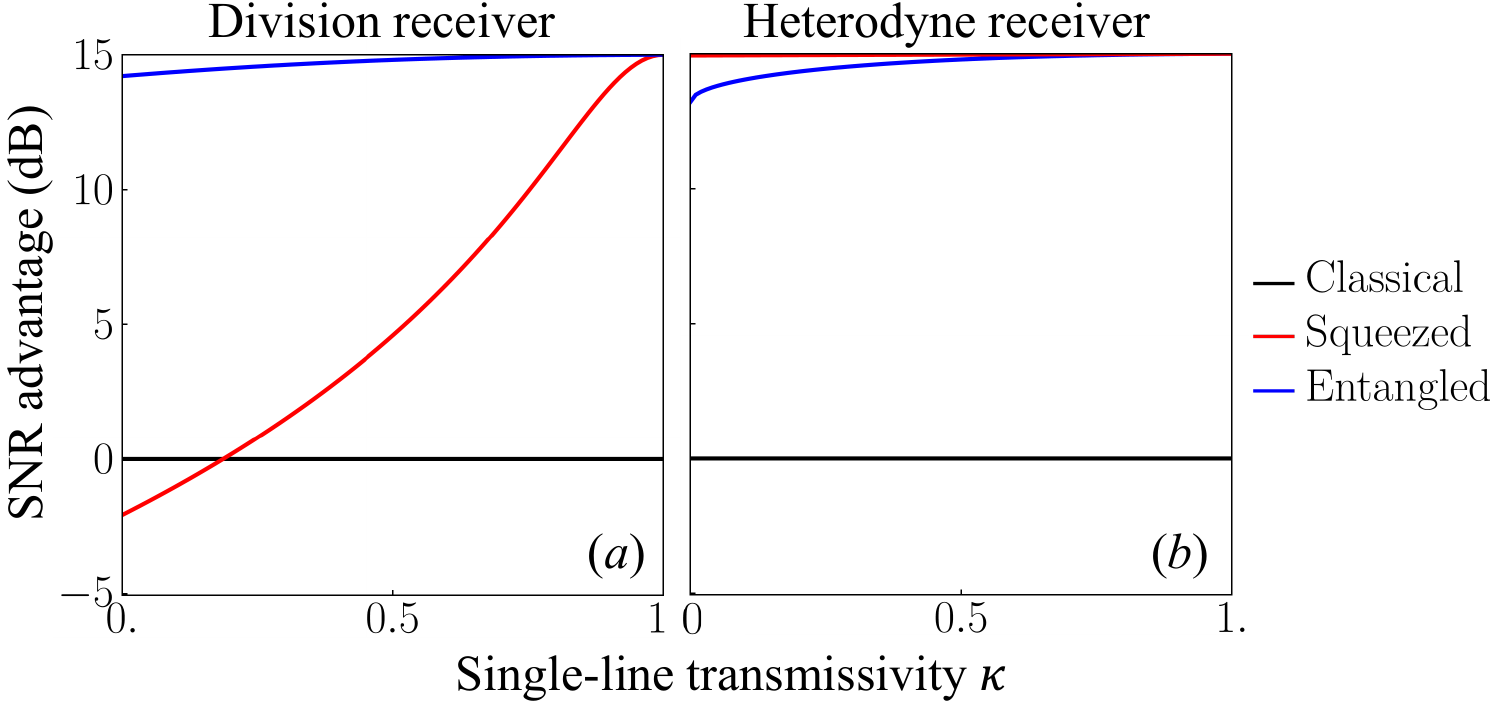}
    \caption{ Quantum advantage in SNR$^2$ over classical DCS. Quantum: $G_A=G_B=15$dB; Classical: $G_A=G_B=1$. $|A|=|B|$, $M=1001$.
    }
    \label{fig:SNRadv_vskappa}
\end{figure}

For heterodyne detection (Fig.~\ref{fig:schematic}a) with intra-comb-line squeezing (Fig.~\ref{fig:schematic}c), the inverse local SNR 
\begin{align} 
&{\rm SNR}_{\rm het}^{-2} \simeq \frac{M}{A^2B^2} \left(\frac{A^2}{G_{B}}  +  \frac{B^2}{G_{A}} \right).
\label{SNR_het_singleline}
\end{align}
For heterodyne detection (Fig.~\ref{fig:schematic}a) with cross-comb-line entanglement (Fig.~\ref{fig:schematic}d), the SNR is identical to Eq.~\eqref{SNR_het_singleline} to the leading order; while inferior to the squeezing performance in higher orders (see Fig.~\ref{fig:SNRadv_vskappa}b).

For the division receiver configuration (Fig.~\ref{fig:schematic}b) and intra-comb-line squeezing (Fig.~\ref{fig:schematic}c)
\begin{align}
{\rm SNR}^{-2}_{\rm div,intra}&
\simeq  \frac{M}{16\kappa A^2B^2}\Big[(3+\kappa)^2(\frac{A^2}{G_A}+\frac{B^2}{G_B})
\nonumber
\\
&\qquad+(1-\kappa)^2(A^2G_B^\prime+B^2G_A^\prime)\Big].
\label{SNR_div_single_line}
\end{align}
We see that the entanglement-mismatching-induced amplified noises proportional to $G_{C}^\prime\equiv\frac{1}{2}(G_{\QZZ{C}}+\frac{1}{G_{\QZZ{C}}})$ for $\QZZ{C}=A,B$ inevitably enter into the variance, even if there is merely a single absorption line.

For the division receiver (Fig.~\ref{fig:schematic}b), cross-comb-line entanglement (Fig.~\ref{fig:schematic}d) yields 
\begin{align}
{\rm SNR}^{-2}_{\rm div,cross}\simeq &
  \frac{M}{16\kappa A^2B^2}\Big[(3+\kappa)^2(\frac{A^2}{G_A}+\frac{B^2}{G_B})
\nonumber
\\
&\qquad\qquad+(1-\kappa)^2(\frac{A^2}{G_B}+\frac{B^2}{G_A})\Big],
\label{SNR_div_single_line_E}
\end{align}
which does not suffer from the entanglement-mismatching-induced amplified noise.

\begin{figure*}
    \centering
    \includegraphics[width=0.8\linewidth]{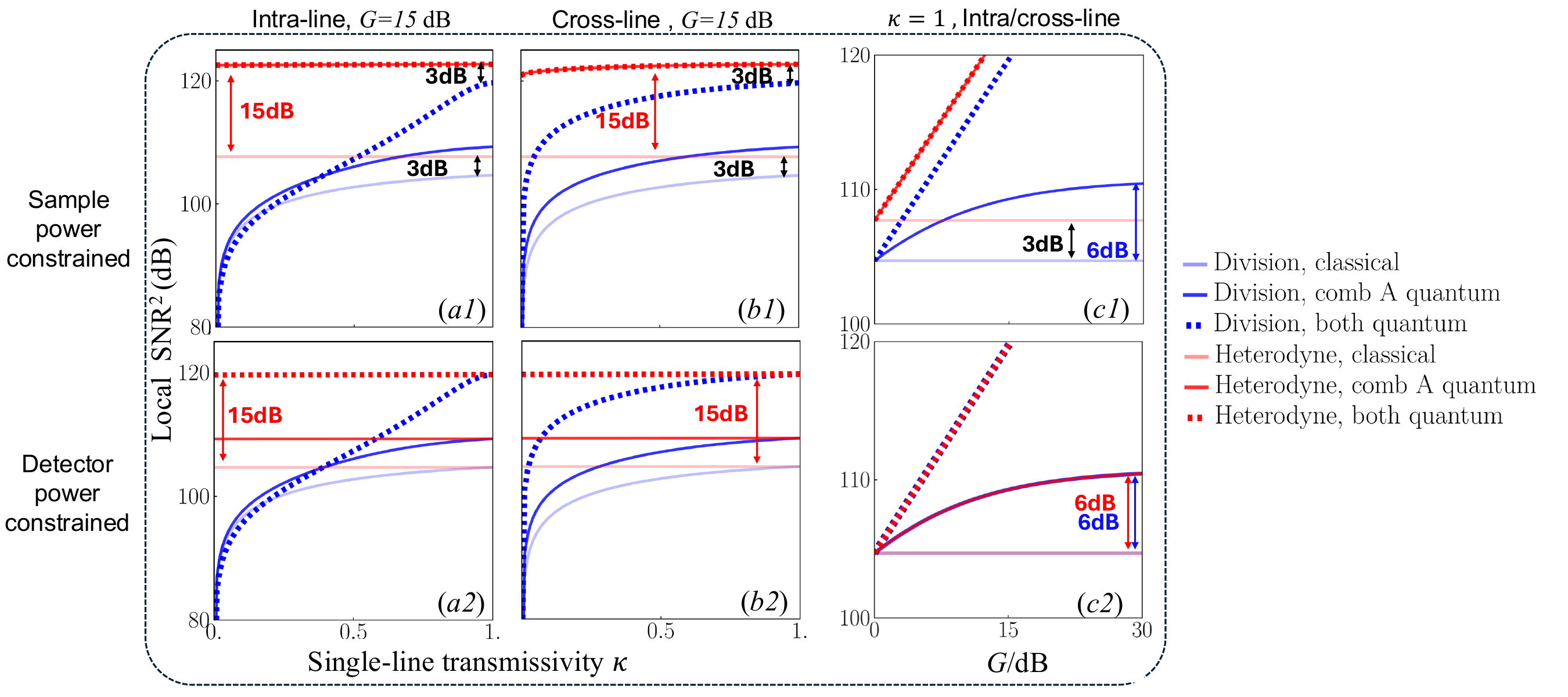}
    \caption{
    Absolute SNR for quantum and classical DCS. 
    We consider SNR of the division receiver (blue) and the heterodyne detection (red). \QZZ{The left and middle columns (a,b) refer to quantum engineering with intra-comb-line squeezing and cross-comb-line entanglement, with $G=15$dB fixed; While in the right column (c) the two are identical due to setting $\kappa=1$.} \hww{ Top row (a1, b1, c1): sample power constrained; Bottom row (a2, b2, c2): detector power constrained. For $G=15$dB in subplots (a)(b), the best quantum advantage over the classical benchmark (dimmed line) is 15dB achieved by the heterodyne with both combs quantum (red dots). In subplots (a1)(b1)(c1) with sample power constrained and both combs quantum (dots), heterodyne detection possesses 3dB advantage over the division receiver. In subplot (c), for all protocols with only comb A quantum (solid) except for heterodyne (red solid) in subplot (c1), there is an upper bound of 6dB advantage over the classical benchmark.}
    $M=1001$, total sample exposure $P=15$mW, carrier wavelength $\lambda=1563$nm. Acquisition time $T=1s$. 
    }
    \label{fig:DivVsHet_SNR}
\end{figure*}

With the SNRs in hand, \QZ{in Fig.~\ref{fig:SNRadv_vskappa}} we evaluate the quantum advantage \QZ{(over the corresponding classical protocols}, using full expressions in \cite{supp}) for the symmetric case of $|A|=|B|$, where both combs are required to be quantum engineered \QZ{in all protocols} to enable quantum advantage beyond 3dB. We observe that three of the four protocols: balanced heterodyne with both intra-comb-line squeezing and cross-comb-line entanglement, and the division receiver with cross-comb-line entanglement, are robust against the single-line loss. In the $M\gg G$ limit, these protocols have negligible decrease of SNR advantage even when $\kappa=0$ and the decay shown in Fig.~\ref{fig:SNRadv_vskappa} is due to finite $M$. In contrast, the division receiver with intra-comb-line squeezing is highly sensitive to loss, due to the entanglement-mismatching-induced amplified noises. When both using the cross-comb-line entanglement, the division receiver yields better SNR advantage than heterodyne, but we note that their corresponding classical limits are different. In fact, later in Fig.~\ref{fig:DivVsHet_SNR} we will show that the heterodyne receiver always yields higher SNR than the division receiver. 


While quantum enhancement against its own classical corresponding protocol is important, the more important metric to optimize is the SNR quantum advantage against the best classical protocol, under certain resource constraints. Below, we consider two types of power resource constraints and evaluate the quantum advantage in terms of the SNR in Fig.~\ref{fig:DivVsHet_SNR}. \hww{We plot three cases for the gain setup, both combs quantum, $G_A=G_B=G$, in dots; only comb A quantum, $G_A=G, G_B=1$, in solid line;  both classical, $G_A=1, G_B=1$, in dimmed line.} The power allocation is optimized to maximize SNR in all cases. In subplots (a) and (b), we see that for $G=15$dB, heterodyne with both quantum combs (red dashed) attain robust 15dB SNR advantage regardless of $\kappa$; \hww{In subplot (a) the division receiver SNRs with both combs quantum (blue dashed) rapidly decay as $\kappa$ drops below unity, due to entanglement-mismatching-induced amplified noise, in contrast in subplot (b) the decaying trend is improved significantly for $\kappa$ not too small, while in both subplots (a) and (b) the division receiver local SNRs decay rapidly as $\kappa\to 0$ due to the signal $\propto\kappa$ dependence giving a $\sqrt{\kappa}$ factor when estimating $\sqrt{\kappa}$ locally $\propto\sqrt{\kappa}$ [see Eq.\eqref{I_mf_mean_main}]. In subplots (a1)(b1), we see heterodyne only requires comb $A$ quantum under the sample power constraint, since its SNR (solid line) achieves the SNR with both combs quantum (dots).}  In subplots (c) given $\kappa=1$, we see that scalable advantage can be achieved for cases with two quantum combs. In contrast, with a single quantum comb, the only case with scalable advantage is heterodyne detection under sample power constraint; while all other cases, e.g. \cite{herman2024squeezed}, is bounded by a constant $6$dB advantage. The details of the comparison can be found in Appendix C. \QZZ{It is worthwhile to mention that under the sample power constraint, heterodyne has a 3dB advantage over division receiver, even in the lossless classical case. This is because here the noises beating with the equally strong signal and LO combs both contribute, while in the heterodyne case only the noises beating with the LO comb mix in. }


\section{Discussions} 
Among all protocols, intra-comb-line squeezing enhanced heterodyne~\cite{shi2023entanglement} enjoys the best loss tolerance, despite experimental challenges. Intra-comb-line squeezing enhanced division receiver~\cite{herman2024squeezed} is most susceptible to loss due to entanglement-mismatching-induced amplified noise. Instead, we propose to enhance division receiver via cross-comb-line entangled comb to regain robustness against loss.
\hw{The effect on quantum advantage of the additive experiment noises, e.g. detector noise and laser RIN, has been analyzed in \cite{shi2023entanglement}.}





\ 

{\em Acknowledgments---}
QZ acknowledges Zheshen Zhang and Scott Diddams for discussions. QZ and HS acknowledge support from NSF (CCF-2240641, OMA-2326746, 2350153), ONR N00014-23-1-2296, AFOSR MURI FA9550-24-1-0349 and DARPA (HR0011-24-9-0362, HR00112490453, D24AC00153-02). This work was partially funded by an unrestricted gift from Google.
QZ proposed and designed the study. HS derived all results and generated all data, with inputs from QZ. QZ and HS wrote the manuscript.

{\em Data availability---}The data that support the findings of
this article are openly available \cite{Github}.

%

\section{Appendix A: Quantum model}
To describe a quantum field $\hat A(t)$, it is convenient to introduce the field operator 
\be 
\hat A(t) =  \int\!\frac{{\rm d}\omega}{2\pi}\,\hat{A}(\omega)e^{-i(\Omega_c+\omega)t},
\label{ES_main}
\ee 
and its frequency-domain annihilation operator 
$
\hat{A}(\omega)=\int_{-\infty}^\infty dt \hat A(t) e^{i\left(\omega+\Omega_c\right) t},
$
where $\Omega_c$ is the carrier frequency. 
The annihilation operator $\hat{A}(\omega)$ satisfies the commutation relation
$
[\hat{A}(\omega),\hat{A}^\dagger(\omega^\prime)]=2\pi \delta\left(\omega-\omega^\prime\right),
$ 
so that the overall field operator satisfies the commutation relation 
$  
[\hat A(t),\hat A^\dagger(t^\prime)]=\delta\left(t-t^\prime\right).
$


For a field with a finite duration $\mathcal{T}=[-T/2,T/2]$, the continuous-time description reduces to a discrete sum,
\be
\hat A(t) = \frac{1}{\sqrt{T}}\sum_\ell \hat{a}_\ell e^{-i2\pi \ell t/T},\mbox{ for $t\in \mathcal{T}$},
\ee
where the modal annihilation operators, 
$
\hat{a}_\ell = \frac{1}{\sqrt{T}}\int_{\mathcal{T}}\!{\rm d}t\,\hat A(t){e^{i2\pi \ell t/T}},
$
satisfy the Kronecker-$\delta$ commutation relation $[\hat a_\ell, \hat a_{\ell^\prime}^\dagger]=\delta_{\ell,\ell^\prime}$.

However, such a discrete set of modes is more than what we need. Moreover, the single index is inconvenient in describing the quantum combs.
As indicated in Eqs.~\eqref{eq_noise_main}, we introduce a double subscript coordinate for the comb field operators 
\bal 
\hat A(t)&=\frac{1}{\sqrt{T}}\sum_{n=-N}^N\sum_{m=-2N}^{2N} e^{-i (n\omega_r+m\Delta\omega_r) t}(A_n\delta_{n,m} + \hat \calA_{n,m}),
\\
\hat B(t)&=\frac{1}{\sqrt{T}}\sum_{n=-N}^N\sum_{m=-2N}^{2N} e^{-i (n\omega_r+m\Delta\omega_r) t}(B_n\delta_{0,m} + \hat \calB_{n,m}), 
\label{eq_combAB_main}
\eal 
where $A_n,B_n$ are dimensionless amplitudes of the comb lines introduced in Eqs.~\eqref{eq_combAB_mean_main_text}, the Kronecker delta $\delta_{j,k}=1$ for $j=k$, otherwise 0, $ \hat \calA_{n,m}$ and $\hat \calB_{n,m}$ are zero-mean quantum noise modes of the input fields in Eq.~\eqref{eq_noise_main}. The number of comb lines $M=2N+1$. 

As shown in Fig.~\ref{fig:schematic_mode_main}, we define the second subscript $m$ to denote the detuning from the absolute frequency $n\omega_r$ in integers of $\Delta\omega_r$ to track the frequency beating between the two combs. For convenience of our analyses, here we assume that each comb line is sharp and therefore only excites the mode at frequency $n \omega_r$. Finite spreading of the comb line $\ll \Delta \omega_r$ will not change the results. Also we assume that $2N\Delta \omega_r<\omega_r$, such that the modes are not overlapping.

We will consider having the comb probing the sample. In terms of the mean field, when two combs go over the sample modeled in Eq.~\eqref{loss_channel}, one can directly obtain the output 
\bal 
\expval{\calL[\hat A(t)]}&=\frac{1}{\sqrt{T}}\sum_{n=-N}^N e^{-i n(\omega_r+\Delta\omega_r) t} \sqrt{\kappa_n } A_n e^{i\theta_n}, 
\\
\expval{\calL[\hat B(t)]}&=\frac{1}{\sqrt{T}}\sum_{n=-N}^N e^{-i n\omega_r t} \sqrt{\kappa_n } B_n e^{i\theta_n},
\label{eq_combAB_mean_loss_main}
\eal 
where we have approximated $\kappa_n=\kappa(n(\omega_r+\Delta\omega_r))\simeq \kappa(n\omega_r)$ and $\theta_n=\theta(n(\omega_r+\Delta\omega_r))\simeq \theta (n\omega_r)$.

As the power change due to quantum squeezing is negligible compared to the comb line mean field power, the quantum comb power,
\begin{align}
P_{\QZZ{C}}=\frac{1}{T}\sum_{n=-N}^N |{\QZZ{C}}_n|^2 \hbar (\Omega_c+n\omega_{\QZZ{C}})\simeq \frac{\hbar \Omega_c}{T}\sum_{n=-N}^N |{\QZZ{C}}_n|^2,
\label{power_AB_main}
\end{align}
where ${\QZZ{C}}=A,B$ and $\omega_A\equiv \omega_r^\prime,\omega_B\equiv \omega_r$ and we have taken the approximation that the carrier frequency $\Omega_c\gg N\omega_r, N\omega_r^\prime$ for simplicity. The more general case is simple but tedious to deal with.

\section{Appendix B: Two-mode squeezing}
Two modes $\hat a_1$ and $\hat a_2$ are in a two-mode squeezed vacuum state (TMSV) if their EPR quadrature variances
\be 
{\rm var}(\hat{q}_1+\hat{q}_2)=1/G, {\rm var}(\hat{p}_1-\hat{p}_2)=1/G.
\label{sqz_var_main}
\ee 
At the same time the anti-squeezing
$
{\rm var}(\hat{q}_1-\hat{q}_2)=G, {\rm var}(\hat{p}_1+\hat{p}_2)=G.
$
Here we have defined the quadrature operators $\hat{q}_k=(\hat a_k+\hat a_k^\dagger)/\sqrt{2}$ and $\hat{p}_k=(\hat a_k-\hat a_k^\dagger)/\sqrt{2}i$. For the TMSV pair, the variance of the complex field operator sum $\hat{a}_1+\hat{a}_2^\dagger$ is entirely suppressed, which is applied in suppressing the noises of Eq.~\eqref{noise_maintext}, \eqref{eq:photocurrent_spectrum_A_main},\eqref{eq:photocurrent_spectrum_B_main}, and
\eqref{noise_maintext_pairing}.

\begin{table*}[t]
\centering
\resizebox{2\columnwidth}{!}{
\begin{tabular}{ ccccccc } 
\toprule[.1em]
\multirow{2}{4em}{Detection scheme} & \multirow{2}{5em}{Quantum comb type} & SNR  &  \multirow{2}{4em}{Loss robust?} & \multirow{2}{5em} {Phase sensitive?}& \multirow{2}{14em}{\# of Q. combs required for \QZ{beyond-constant} Q. advantage} & {Previous works}\\
&&&&&&\\
\midrule[.1em]
heterodyne & squeezed  & Eq.~\eqref{SNR_het_singleline} & robust & yes& 1 (sample) or 2 (detector)& Theory in Ref.~\cite{shi2023entanglement} \\
heterodyne & entangled  & Eq.~\eqref{SNR_het_singleline} & robust & yes &  1 (sample) or 2 (detector)& Exp. in Ref.~\cite{hariri2025entangled}\\
division& squeezed& Eq.~\eqref{SNR_div_single_line} & not robust & no & 2 for both cases& Exp. in Ref.~\cite{herman2024squeezed}, with only 1 Q. comb\\
division& entangled & Eq.~\eqref{SNR_div_single_line_E}& robust & no & 2 for both cases& Theory in this work\\
\bottomrule[.1em]
\end{tabular}
}
\caption{Summary of the comparison between four protocols. 
The number of required quantum combs depends on whether the constraint is on sample power (indicated by `sample') or detector power (indicated by `power').  Q. is short for quantum, and Exp. is for experiment. 
\label{table_summary}
}
\end{table*}


\section{Appendix C: Power-constrained SNR analyses}
Here we provide details of the sample power constraint and detector power constraint. First, we assume both $A$ and $B$ can be squeezed and their maximum squeezing gains are identical, then we discuss the special case where only one comb can be squeezed.

\subsection{Sample power constraint} First, we consider a scenario where the probing power on the sample is limited to $P$, such as in bio-sensing of tissues~\cite{casacio2021quantum,shi2023entanglement}. 

For heterodyne detection, as only the signal passes through the sample, the power constraint only applies to the signal comb amplitude square, $|A|^2=PT/M\hbar\Omega_c$, where \QZZ{$P$ is the power constraint,} $T$ is the duration of probing and $\Omega_c$ is the carrier frequency; while the LO amplitude square $|B|^2$ can be arbitrarily large. Indeed, it is well-known that the maximum SNR is achieved at the strong LO limit $|B|^2\gg1$ in the classical case, and the SNR
\be 
{\rm SNR}_{\rm C}^{\star2}\simeq {PT}/{M^2\hbar\Omega_c},
\label{eq:SNR_C_app}
\ee 
independent of $\kappa$. \hw{Comparing with the well-known result Eq.~(2) in \cite{newbury2010sensitivity}, our classical SNR exactly recovers it as ${\rm SNR}_{\rm C}^{\star 2}=1/\sigma_H^2$ given $a_{\rm shot}=\hbar\Omega_c$, $N_d=1$, $\epsilon=1$, ignoring the Gaussian envelop factor 0.8.} At the same strong LO limit, only the signal comb needs to be quantum engineered. With gain $G_A=G$, the quantum DCS protocol has
\be 
{\rm SNR}_{\rm het}^{\star2}\simeq  G \times {\rm SNR}_{\rm C}^{\star2},
\label{eq:SNR_het_quantum_app}
\ee 
regardless of the loss induced from a single-comb line absorption. \hw{We note that the quantum advantage of single-mode squeezing is limited to 3dB~\cite{collett1987quantum}, because heterodyne picks up an extra vacuum noise at the image frequency than homodyne; in contrast, here we achieve an advantage scalable with $G$, as a benefit of two-mode squeezing between the signal and image frequencies~\cite{collett1987quantum,zhang2021two}, in a similar spirit of twin beam heterodyne detection~\cite{d1997equivalence}.
On the other hand, squeezing does not resolve the 3dB penalty on heterodyne compared with homodyne, which requires a two-carrier input~\cite{zhang2021two}.  }

In the division receiver case, the two input combs mixes and then jointly impinge on the sample. Therefore, the power constraint leads to $(|A|^2+|B|^2)/2=PT/M\hbar\Omega_c$.  
To achieve a substantial quantum advantage over the classical performance, a division receiver requires two quantum combs in this case. Here, we assume $G_A=G_B=G$, which achieves the maximum SNR at $A^2=B^2$ and
\be 
{\rm SNR}^{\star2}_{\rm div}\simeq  8\kappa \left[\frac{\left(3+\kappa\right)^2}{G}+\left(1-\kappa\right)^2\frac{G+1/G}{2}\right]^{-1} {\rm SNR}_{\rm C}^{\star2}.
\label{eq:SNR_div_quantum}
\ee


\QZ{The classical case can be directly obtained by setting $G=1$ in Eq.~\eqref{eq:SNR_div_quantum} above, leading to ${\rm SNR}^{\star2}_{\rm div}/{\rm SNR}_{\rm C}^{\star2}\simeq {4\kappa }/{[4+(1+\kappa)^2]}$.}
Even at the lossless case of $\kappa=1$, we have the division receiver a factor of two worse than the heterodyne in the classical case, ${\rm SNR}^{\star2}_{\rm div}\simeq {\rm SNR}_{\rm C}^{\star2}/2$. 
This is because here the noises beating with the equally strong signal and LO combs both contribute, while in the heterodyne case only the noises beating with the LO comb mix in. \QZ{Such a gap extends to the quantum cases. This explains the 3dB gap between division receiver and heterodyne receiver in the sample power constraint cases of Fig.~\ref{fig:DivVsHet_SNR} (a1) and (b1).}

\subsection{Detector power constraint}
On the other hand, if only the detector saturation, instead of the power exposure on the sample, is concerned, then the total power is constrained: $(|A|^2+|B|^2)/2=PT/M\hbar\Omega_c$. 

In this case, both signal and LO combs are required to be quantum engineered for both heterodyne and division receivers for substantial quantum advantages. We can also show that for both receivers, under a maximum (two-mode) squeezing gain limit, the optimal strategy is to have symmetric quantum combs, with $G_A=G_B=G$ and $|A|^2=|B|^2$. Under the above settings, the division receiver has the same performance as in the case of sample power constraint, given by Eq.~\eqref{eq:SNR_div_quantum}. 
For heterodyne detection, the optimal SNR is 
\be 
{\rm SNR}_{{\rm het}, |A|=|B|}^{\star 2}=\frac{G}{2} {\rm SNR}_{\rm C}^{\star2},
\label{eq:SNRhet_detectorConstr_app}
\ee
with the classical performance given by $G=1$ as
$
{\rm SNR}_{\rm het}^{\star2}\simeq {\rm SNR}_{\rm C}^{\star2}/2.
$ 
This is due to the absence of an infinitely strong LO comb.

\QZ{Under both power constraints, the heterodyne receiver overwhelms the division receiver in the high absorption region $\kappa\sim 0$, since the SNR is defined using Fisher information which diminishes at the lossy limit $\kappa\to 0$ for division receiver (see Eq.~\eqref{eq:SNR_div_quantum}). Such a gap disappears when one consider a global SNR motivated by hypothesis testing instead of parameter estimation, as we discuss in \cite{supp}.}


%

\hw{
\subsection{Only one comb squeezed} 
We take comb A squeezed as example, $G_A=G$, $G_B=1$. Consider the limit $M\gg G$. At the lossless limit $\kappa\to 1$, the division receiver with both intra-comb-line squeezing and cross-comb-line entanglement have the same 
$
{\rm SNR}^2_{div,\kappa\to 1}\simeq \frac{A^2 B^2 G}{A^2 M+B^2 M G}\,.
$
For both power constraints, the division receiver has the same optimal power allocation $|A|^2=\frac{\sqrt{G}}{1+\sqrt{G}}\frac{2PT}{M\hbar\Omega_c}$, which achieves 
$
{\rm SNR}^2_{div,\kappa\to 1, opt}=\frac{2 P G}{M \left(\sqrt{G}+1\right){}^2}\,.
\label{eq:SNR_div_lossless_powewopt_app}
$
At strong squeezing limit $G\to \infty$, it gives a 6dB advantage over classical $G=1$.
On the other hand, the heterodyne detection is independent on the single-line absorption $\kappa$, the SNR is given by Eq.~\eqref{SNR_het_singleline}. For sample power constrained case, the power constraint $|A|^2=\frac{PT}{M\hbar\Omega_c}$ gives 
$
{\rm SNR}^2_{\rm het,sample, opt}=\frac{G P}{M}\,,
$
for both intra-comb-line squeezing and cross-comb-line entanglement. It keeps the scalable advantage intact. For detector power constrained case, the heterodyne detection has the optimal power allocation $|A|^2=\frac{1}{1+\sqrt{G}}\frac{2PT}{M\hbar\Omega_c}$, which achieves 
$
{\rm SNR}^2_{\rm het,detector, opt}=\frac{2 G P}{\left(\sqrt{G}+1\right)^2 M}\,.
$
We see that it is identical to the $\kappa\to 1$ limit of the division receiver. Hence in Fig.~\ref{fig:DivVsHet_SNR}(c), for the one-comb-squeezed cases (dot-dashed), we observe that only the heterodyne under the sample power constraint achieves advantage scalable with $G$, while all other cases are limited by a constant 6dB advantage. 
}

\hw{
\section{Appendix D: Multiple absorption lines} 
The single-absorption-line case can be easily generalized to $\gamma M$-absorption-line case. For simplicity, we assume uniform transmissivity $\kappa$ for the $\gamma M$ lines, and $\gamma$ is small so that the $\gamma M$ absorption lines locates at the same side of the carrier. In this case, SNRs of division receiver are given by Eqs.~\eqref{eq:SNR_div_intra_gammaM_supp}\eqref{eq:SNR_div_cross_gammaM_supp} in \cite{supp} for intra-comb-line squeezing and cross-comb-line entanglement respectively. SNRs of heterodyne detection are given by Eqs.~\eqref{eq:SNR_het_intra_gammaM_supp}\eqref{eq:SNR_het_cross_gammaM_supp}. Take the limit $\gamma G\ll 1-\gamma$, we recover the single-absorption-line results. The loss robustness degrades as $\gamma$ increases towards unity.
}

\newpage 
\ 
\newpage

\renewcommand*{\thesection}{SM  \arabic{section}}
\renewcommand*{\thesubsection}{SM \arabic{section}.\arabic{subsection}}
\renewcommand*{\thesubsubsection}{SM \arabic{section}.\arabic{subsection}.\arabic{subsubsection}}
\addto\captionsenglish{\renewcommand{\figurename}{Supplementary Figure}}
\renewcommand{\tablename}{Supplementary Table}

\renewcommand{\thefigure}{S\arabic{figure}}
\renewcommand{\theequation}{S\arabic{equation}}
\setcounter{section}{0}
\setcounter{equation}{0}
\setcounter{figure}{0}
\setcounter{table}{0}

\begin{widetext}
\tableofcontents

\newpage 
\begin{center}
{\Large {\bf
Supplementary Materials:\\
Theory of quantum comb enhanced interferometry
}
}
\end{center}
\end{widetext}

\hw{
\section{Index of the derivations of the main results in main text}
}
\begin{table}[h!]
  \renewcommand{\arraystretch}{1.6}
    \centering
    \begin{tabular}{c| c}
    \hline
    \hline
    Formula in main text & Derivation in this SM
    \\
    \hline
    Eq.~\eqref{d_m_het}
    &
    Eq.~\eqref{eq:d_m_het_supp}
    \\
    Eq.~\eqref{noise_maintext} 
    &
    Eq.~\eqref{eq:d_noise_supp}
    \\
    Eq.~\eqref{I_mf_mean_main}
    &
    Eq.~\eqref{eq:I_mf_mean_supp}
    \\
    Eqs.~\eqref{eq:photocurrent_spectrum_A_main}\eqref{eq:photocurrent_spectrum_B_main}
    &
    Eqs.~\eqref{eq:photocurrent_spectrum_A_supp}\eqref{eq:photocurrent_spectrum_B_supp}
    \\
    Eq.~\eqref{IA_IB_division_sum_main}
        &
    Eq.~\eqref{IA_IB_division_sum}
    \\
    Eq.~\eqref{SNR_het_singleline} 
    &   Eq.~\eqref{eq:var_het_intra_singleline_supp}
\\
    Eq.~\eqref{SNR_div_single_line} & Eq.~\eqref{eq:var_div_intra_singleline_supp} 
    \\
Eq.~\eqref{SNR_div_single_line_E}
& Eq.~\eqref{eq:var_div_cross_singleline_supp}
    \end{tabular}
    \caption{ 
    \hw{The location in this Supplementary Materials (SM) of the derivations of the main results in main text.
    }
    }
    \label{tab:main_formulas}
\end{table}

\section{Theory framework}
\label{sec:framework_supp}
To describe the field, we use the field annihilation operator $\hat{A}$, which satisfies the commutation relation
\be 
[\hat{A}(\omega),\hat{A}^\dagger(\omega^\prime)]=2\pi \delta\left(\omega-\omega^\prime\right),
\ee 
in spectral domain,
and
\be  
[\hat A(t),\hat A^\dagger(t^\prime)]=\delta\left(t-t^\prime\right)\,,
\ee
in time domain. The total mean photon number of the pulse can be calculated as
$
\int_{-\infty}^\infty dt \expval{\hat A^\dagger(t)\hat A(t)}= 
\frac{1}{2\pi}\int_{-\infty}^{\infty} d\omega  \expval{\hat{A}^\dagger(\omega)\hat{A}(\omega)},
$
while the energy is
\be 
E=\frac{1}{2\pi}\int_{-\infty}^{\infty} d\omega \hbar(\omega+\Omega_c)  \expval{\hat{A}^\dagger(\omega)\hat{A}(\omega)}.
\label{energy_def}
\ee 
Here $\Omega_c$ is the carrier frequency. For simplicity, we set $\Omega_c=0$ in our analysis, except for the energy calculations where we assume $\omega+\Omega_c\simeq \Omega_c$.

\begin{figure*}
    \centering
    \includegraphics[width=\linewidth]{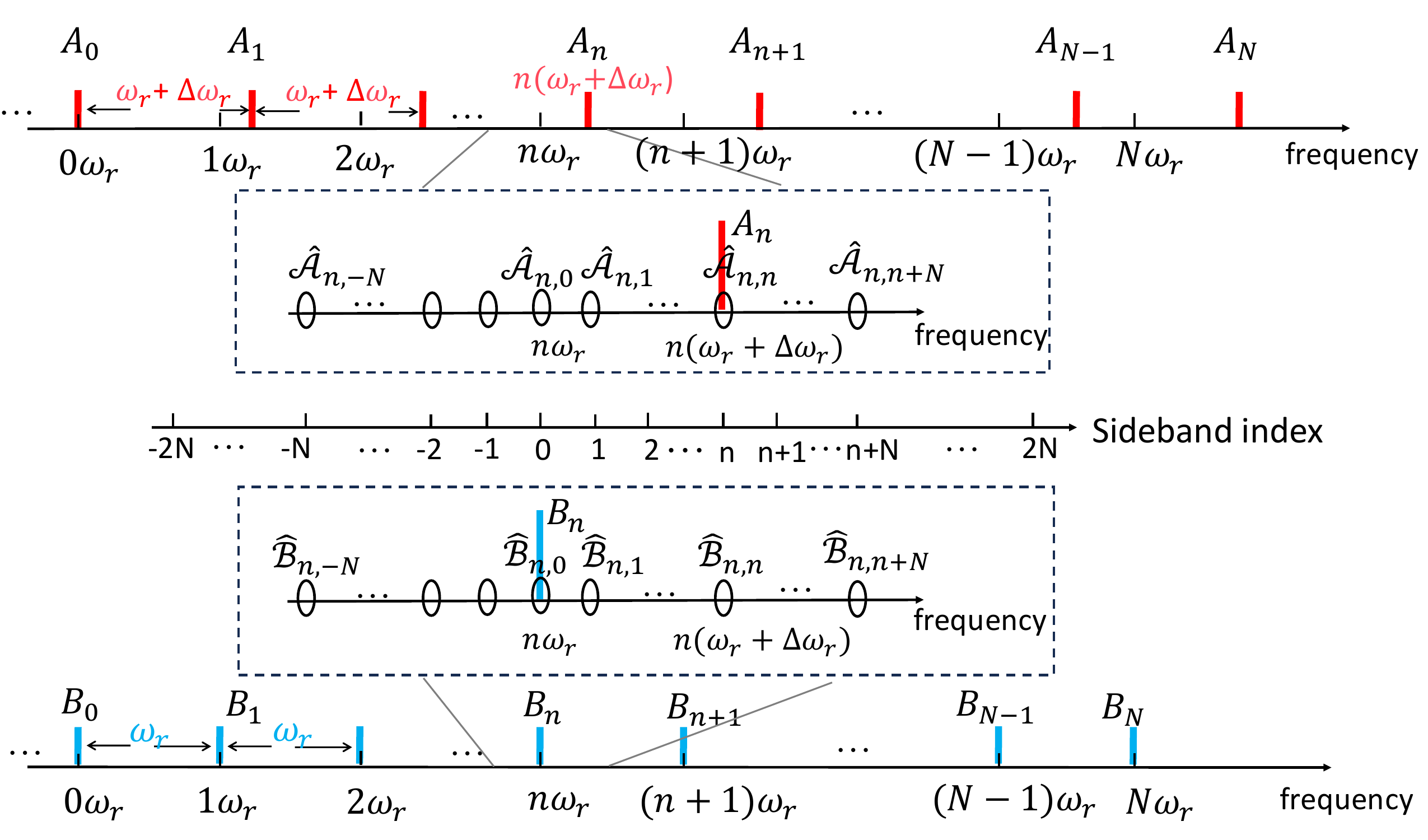}
    \caption{ Schematic of the mode definitions. The comb lines are denoted by $A_n,B_n$, located at frequency $n(\omega_r+\Delta \omega_r)$ and $n\omega_r$ respectively, for $-N\le n\le N$. For the noise modes $\hat \calA_{n,m}, \hat \calB_{n,m}$, the first subscript $n$ indexes the comb line, while the second subscript $m$ indexes the detuning of the sideband noise mode from the line. \hw{To contain all relevant noise modes including both self-beating and cross-beating terms, $\min\{n-N,-N\}\le m\le \max\{n+N,N\}$ at the sideband of the $n$th line. For simplicity, we allow $-2N\le m\le 2N$.} }
    \label{fig:schematic_mode}
\end{figure*}

In dual-comb spectroscopy, only a discrete set of modes are relevant for the frequency beating in the readout photocurrent. Consider $N$ pairs of lines around the carrier frequency, the total number of comb lines is $M=2N+1$. Each line has a sideband containing $N$ pairs of noise modes that will also beat with the lines. To describe the $2M^2$ noise modes per comb, we introduce a double subscript coordinate for the two comb field operators 
\bal 
\hat A(t)&=\frac{1}{\sqrt{T}}\sum_{n=-N}^N\sum_{m=-2N}^{2N} e^{-i (n\omega_r+m\Delta\omega_r) t}(A_n\delta_{n,m} + \hat \calA_{n,m})
\\
\hat B(t)&=\frac{1}{\sqrt{T}}\sum_{n=-N}^N\sum_{m=-2N}^{2N} e^{-i (n\omega_r+m\Delta\omega_r) t}(B_n\delta_{0,m} + \hat \calB_{n,m}) 
\label{eq_combAB}
\eal 
where $T$ is the acquisition time, $A_n,B_n$ are dimensionless amplitudes of the comb lines, the Kronecker delta $\delta_{j,k}=1$ for $j=k$, otherwise 0, $ \hat \calA_{n,m}$ and $\hat \calB_{n,m}$ are zero-mean quantum noise modes of the input fields. \hw{Note that here we include all $m\in[-2N,2N]$ noises with redundancy, the number of relevant noise modes is only $2M^2$ per comb: for $A\hat \calA$ and $B\hat \calB$ self beating, the relevant noises are $\sum_{m=n-N}^{n+N}\hat \calA_{n,m}$ for comb $A$ and $\sum_{m=-N}^{N}\hat \calB_{n,m}$ for comb $B$; for $A\hat \calB$ and $B\hat \calA$ cross beating, the relevant noises are $\sum_{m=-N}^{N}\hat \calA_{n,m}$ for $A$ and $\sum_{m=n-N}^{n+N}\hat \calB_{n,m}$ for $B$.} As shown in Fig.~\ref{fig:schematic_mode}, we define the second subscript $m$ to denote the detuning from the absolute frequency $n\omega_r$ instead of the comb line frequency $n(\omega_r+\Delta\omega_r)$, because it is easier to track the frequency beating for cross-beating terms. For convenience of our analyses, here we assume that each comb line is sharp and therefore only excites the mode at frequency $n \omega_r$. Finite spreading of the comb line $\ll \Delta \omega_r$ will not change the results. Also we assume that $N\Delta \omega_r<\omega_r$, such that the modes are not overlapping. 

The quantum description of Eq.~\eqref{eq_combAB} yields the mean
\bal 
\expval{\hat A(t)}&=\frac{1}{\sqrt{T}}\sum_{n=-N}^N\sum_{m=-2N}^{2N} e^{-i (n\omega_r+m\Delta\omega_r) t} A_n\delta_{n,m} 
\\
&=\frac{1}{\sqrt{T}}\sum_{n=-N}^N e^{-i n(\omega_r+\Delta\omega_r) t} A_n 
\\
\expval{\hat B(t)}&=\frac{1}{\sqrt{T}}\sum_{n=-N}^N\sum_{m=-2N}^{2N} e^{-i (n\omega_r+m\Delta\omega_r) t} B_n\delta_{0,m}  
\\
&=\frac{1}{\sqrt{T}}\sum_{n=-N}^N e^{-i n\omega_r t} B_n.
\label{eq_combAB_mean}
\eal 
which agrees with the classical description of comb lines.

With the comb line amplitudes defined, we can evaluate the power for later usage,
\begin{subequations}
\begin{align}
P_A&=\frac{1}{T}\sum_{n=-N}^N |A_n|^2 \hbar [\Omega_c+n(\omega_r+\Delta\omega_r)]
\nonumber\\
&\simeq \frac{\hbar \Omega_c}{T}\sum_{n=-N}^N |A_n|^2,
\\
P_B&=\frac{1}{T}\sum_{n=-N}^N |B_n|^2 \hbar [\Omega_c+n\omega_r]\simeq \frac{\hbar \Omega_c}{T}\sum_{n=-N}^N |B_n|^2,
\end{align}
\label{power_AB}
\end{subequations}
where we have taken the approximation that $\Omega_c\gg N\omega_r, N\Delta \omega_r$ for simplicity. We note that the power change due to quantum squeezing is negligible compared to the comb line mean field power.

Now consider the comb probing the sample. \hw{First we reiterate the sample model Eq.~\eqref{loss_channel} of main text:
\be 
X_n^\prime= \sqrt{\kappa_n}e^{i\theta_n}X_n, \, \hat \calX^\prime_{n,m}= \sqrt{\kappa_n}e^{i\theta_n}\hat \calX_{n,m}+{\cdots},
\label{loss_channel_app}
\ee 
with `$\cdots$' denoting the loss-induced noise.
In this paper we consider two types of probing schemes: ``combine then pass through sample'' as elaborated in \ref{sec:combine_pass}, and ``pass through sample then interfere'' as elaborated in \ref{sec:pass_combine}.
For the former, in terms of the mean field, the outputs after the two combs travel through the sample are
\bal 
\expval{\calL[\hat A(t)]}&=\frac{1}{\sqrt{T}}\sum_{n=-N}^N e^{-i n(\omega_r+\Delta\omega_r) t} \sqrt{\kappa_n } A_n e^{i\theta_n} 
\\
\expval{\calL[\hat B(t)]}&=\frac{1}{\sqrt{T}}\sum_{n=-N}^N e^{-i n\omega_r t} \sqrt{\kappa_n } B_n e^{i\theta_n},
\label{eq_combAB_mean_loss}
\eal 
where we have approximated $\kappa(n(\omega_r+\Delta\omega_r))\simeq \kappa(n\omega_r)=\kappa_n$ and $\theta(n(\omega_r+\Delta\omega_r))\simeq \theta (n\omega_r)=\theta_n$. For the latter, comb $A$ is subject to Eq.~\eqref{eq_combAB_mean_loss} while the local oscillator comb $B$ does not change from Eq.~\eqref{eq_combAB_mean} since it does not travel through the sample. For the loss-induced noise `$\ldots$', we consider it as $\sqrt{1-\kappa_n}\hat \calE_n$, contributed by an environment mode $\calE_n$ in thermal state with mean photon number
\be 
E=\frac{1}{e^{\hbar \Omega_c/k_Bt}-1}\,,
\ee 
with transmissivity $1-\kappa_n$,} where $k_B$ is the Boltzmann constant, $t$ is temperature. Here we approximate $N\omega_r, N\Delta\omega_r\ll \Omega_c$.


\hw{Finally, let us consider the detection. We cover two mainstream detection schemes: division information processing (division receiver), and subtraction data processing (heterodyne detection), which will be elaborated in \ref{sec:combine_pass} and \ref{sec:pass_combine}. Here we briefly introduce the mechanism of squeezing and its advantage in the detection SNR.} The variance of the photocurrent readouts contains complex quadratures $\hat Q_{n,m}$ for the $\cos(m\Delta \omega_r t)$ component and $\hat P_{n,m}$ for the $\sin(m\Delta \omega_r t)$ component, in the form of $\hat a_1\pm \hat a_2^\dagger$, as shown later in Eqs.~\eqref{eq:photocurrent_spectrum_A_supp}\eqref{eq:photocurrent_spectrum_B_supp}, which agrees with the known conclusions of heterodyne detection~\cite{caves1985new}. In a quantum comb, we will engineer the squeezing of the combs to squeeze the photocurrent readout variance, in a way that the comb noise modes are paired and each pair is in a two-mode squeezed vacuum state. We consider two pairing structures of the noise modes. 
One is the `intra-line squeezing', where the squeezed pairs are centered around each line, as shown in Figure~\ref{fig:schematic}(c) in maintext:
\be 
\hat Q_{n,m}^{\rm intra}=\hat \calA_{n,m}+\hat \calA_{n,-m}^\dagger, \hat P_{n,m}^{\rm intra}=(\hat \calA_{n,m}-\hat \calA_{n,-m}^\dagger)/i\,.
\label{eq:squeezed_quad_supp}
\ee  
The other is the `cross-line entanglement', where the squeezed pairs are centered around the carrier frequency, as shown in Figure~\ref{fig:schematic}(d) in maintext:
\be 
\hat Q_{n,m}^{\rm cross}=\hat \calA_{n,m}+\hat \calA_{-n,-m}^\dagger, \hat P_{n,m}^{\rm cross}=(\hat \calA_{n,m}-\hat \calA_{-n,-m}^\dagger)/i\,.
\label{eq:entangled_quad_supp}
\ee  
Note that these quadratures are complex-valued, and both real and imaginary parts can carry the signal since the signal can have an unknown phase, thus we define $\text{var} \hat X\equiv \expval{(\Re\Delta\hat X)^2}+\expval{(\Im\Delta\hat X)^2}$ for any such complex-valued operator $\hat X$, where $\Delta\hat X\equiv \hat X-\expval{\hat X}$. For $\hat X\propto \hat a_1\pm \hat a_2^\dagger$, we have $[\Re\hat X,\Im\hat X]=0$ and $\text{var} \hat X=\expval{\Delta X^\dagger \Delta X}$. 
Under such definition, the vacuum noise equals unity, i.e. $\expval{\hat Q_{n,m}^\dagger \hat Q_{n,m}}=\expval{\hat P_{n,m}^\dagger \hat P_{n,m}}=1$ for vacuum state. 
For intra-line squeezing,
\be 
{\rm var}(\hat Q_{n,m}^{\rm intra})=1/G_n\,,
\label{sqz_var_intra_supp}
\ee 
and antisqueezing
\be 
{\rm var}(\hat P_{n,m}^{\rm intra})=G_n\,,
\label{antisqz_var_intra_supp}
\ee 
For cross-line entanglement,
\be 
{\rm var}\hat Q_{n,m}^{\rm cross}=1/G_n\,,
\label{sqz_var_cross_supp}
\ee 
and antisqueezing
\be 
{\rm var}\hat P_{n,m}^{\rm cross}=G_n\,.
\label{antisqz_var_cross_supp}
\ee 
We have omitted the heterogeneous squeezing within the sideband of each line, i.e. we assume $G_{n,m}=G_n$.
In this work, all quantum engineering protocols are based on two-mode squeezing under the above two pairing setups.

\subsection{SNR definitions}
Consider the estimation of the parameter $\lambda$ from measurement results sampled from a Gaussian distribution of mean $\mu(\lambda)$ and variance $\sigma^2$. From $S$ sampling of the random number, the estimation error is bounded by the Cramer-Rao bound
$
\delta \lambda^2\ge 1/(S \calF_\lambda),
$ 
where the Fisher information $\calF_\lambda=(\partial_\lambda \mu)^2/\sigma^2$. Therefore, we define the SNR as
\be 
{\rm SNR}^2=(\partial_\lambda \mu)^2/\sigma^2.
\label{eq:SNR_point}
\ee 
In spectroscopy, we can evaluate the mean and variance of the measurement results, as a function of $\sqrt{\kappa}$. Then we can evaluate the SNR for estimating the absorption $\sqrt{\kappa}$. However, the SNR defined from Fisher information only captures performance of local parameter estimation---the parameter $\sqrt{\kappa}$ is known to be close to a prior value that the Fisher information is evaluated at. As an alternative approach, we also consider a global version of SNR---taking the finite difference version of the derivative $\partial_\lambda \mu$ and define
\be 
\overline{\rm SNR}^2=\frac{(\mu|_{\kappa}-\mu|_{\kappa=1})^2}{\sigma^2}.
\label{eq:SNR_global}
\ee 
Such a definition applies to scenarios such as hypothesis testing between two possible absorption values $\kappa$ and unity.

We observe that the local SNR of division receiver in Eq.~\eqref{eq:SNR_div_quantum} in the main text with respect to $\sqrt{\kappa}$ diminishes at the lossy limit $\kappa\to 0$. This is because here the photocurrent readout at the sample arm is modulated by $\kappa$, as both of the two beating combs pass through the sample, which can be verified by Eq.~\eqref{I_mf_mean_main}.

Meanwhile, we observe that the local SNR of heterodyne receiver remains nonzero at the lossy limit $\kappa\to 0$, in sharp contrast to the division receiver. This is because here the heterodyne photocurrent readout is modulated by $\sqrt{\kappa}$, not $\kappa$ in the division receiver case, as only one of the two beating combs pass through the sample, which can be verified by Eq.~\eqref{I_mf_mean_het}.

By comparing Eqs.~\eqref{eq:SNR_point} and \eqref{eq:SNR_global}, we see that the local and global SNRs are connected by
\be 
\overline{\rm SNR}^2=\frac{(\mu|_{\kappa}-\mu|_{\kappa=1})^2}{(\partial_{\sqrt{\kappa}} \mu)^2}{\rm SNR}^2\equiv c_{l\to g}(\kappa){\rm SNR}^2\,,
\ee
where we define the local-to-global coefficient $c_{l\to g}(\kappa)\equiv \frac{(\mu|_{\kappa}-\mu|_{\kappa=1})^2}{(\partial_{\sqrt{\kappa}} \mu)^2}$. Given the single-line absorption spectrum in Fig.~\ref{fig:schematic}e in the main text, for division receiver, $\mu(\kappa)=\frac{1+\kappa}{2}$, thus $c_{l\to g}^{\rm div}(\kappa)=\frac{(\kappa-1)^2}{4\kappa}$ ; for heterodyne receiver, $\mu(\kappa)=(\sqrt{\kappa}+1)AB$, thus $c_{l\to g}^{\rm het}(\kappa)=(\sqrt{\kappa}-1)^2$. At the limit of $\kappa\to 1$, $c_{l\to g}^{\rm div}(\kappa)\simeq c_{l\to g}^{\rm het}(\kappa)$.

\begin{figure}
    \centering
    \includegraphics[width=\linewidth]{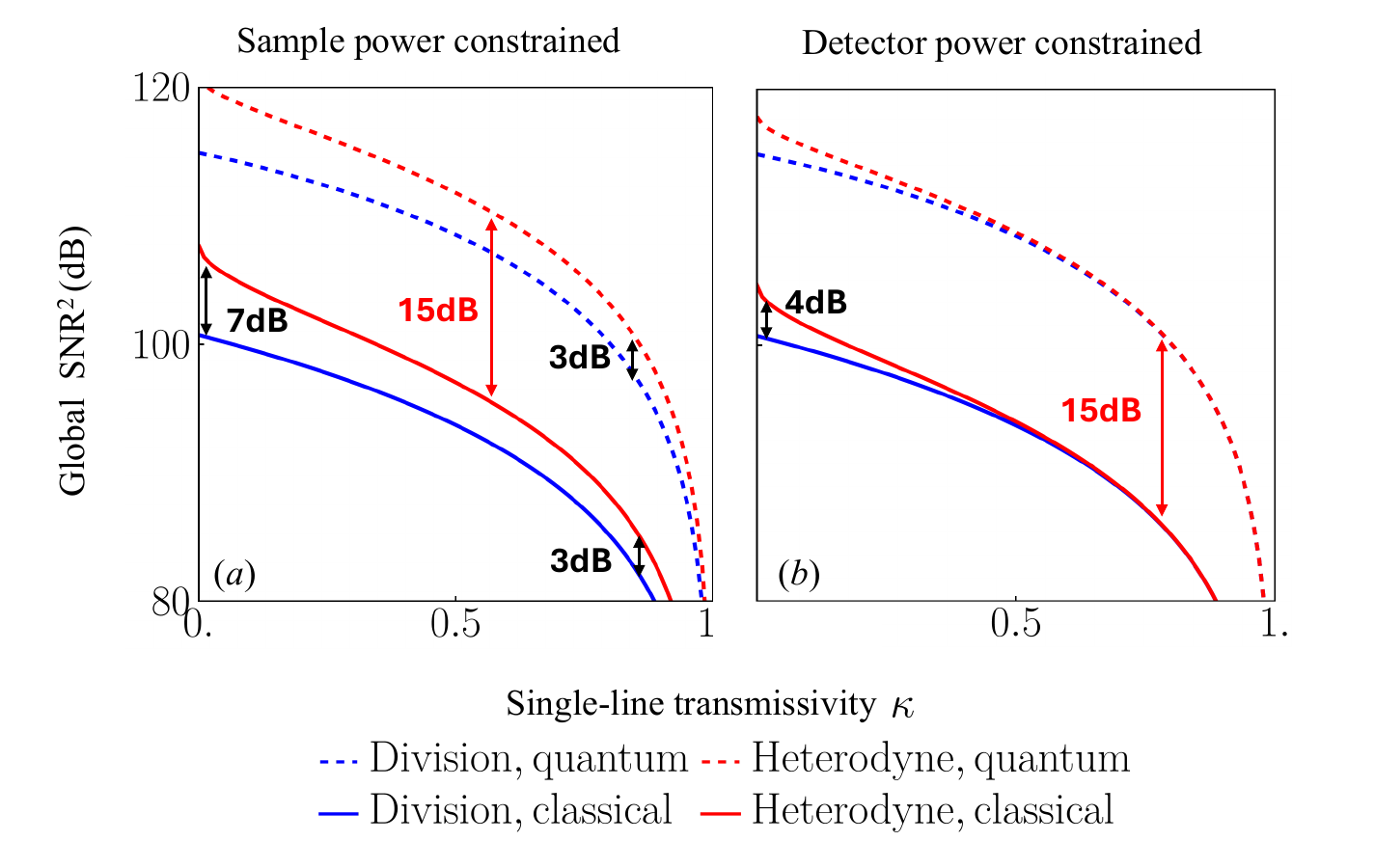}
    \caption{ Absolute SNR for cross-comb-line entangled combs in DCS. We consider the global SNR (Eq.~\eqref{eq:SNR_global}) of the division receiver (blue) and the heterodyne detection (red). The power allocation is optimized to maximize SNR in all cases. Quantum gains setup: $G_A=G_B=15$dB with for all. For all the scenarios, we provide the classical benchmark with $G_A=G_B=1$ (dashed) for reference. $M=1001$, total sample power exposure $P=15$mW, carrier wavelength $\lambda=1563$nm. Acquisition time $T=1s$. }
    \label{fig:DivVsHet_SNR_global}
\end{figure}

\hw{In the maintext we have demonstrated and discussed the numerical evaluation results of local SNR. Here, we focus on the global SNR, which is evaluated in Fig.~\ref{fig:DivVsHet_SNR_global}.} Contrary to the local SNR, the global SNR is always nonzero at $\kappa\to 0$ while it decays to zero as $\kappa\to 1$. Still, we observe the heterodyne receiver is more robust to the loss than the division receiver, even both using the cross-comb-line entanglement which is loss-robust per se. 
At $\kappa= 0$, the advantage of 7dB in the global SNR under sample power constraint (and 4dB under detector power constraint) of heterodyne over division for the classical cases can be predicted by comparing the SNR formulas $\overline{\rm SNR}^2_{\rm het,classical}(\kappa= 0)={1}/{[M(1/A^2+1/B^2)]}$, $\overline {\rm SNR}^2_{\rm div,classical}(\kappa= 0)={2}/{[5M(1/A^2+1/B^2)]}$. Under sample power constraint, the heterodyne receiver is optimized at $A^2=PT/M\hbar\Omega_c$, $B^2\to \infty$, the division receiver is optimized at $A^2=B^2=PT/M\hbar\Omega_c$, then the SNR difference is doubled to around 7dB; Under detector power constraint, both receivers have $A^2=B^2=PT/M\hbar\Omega_c$, then the SNR difference is $5/2\simeq 4$dB for the classical performance. On the other hand, at $\kappa\to 1$, the 15dB quantum advantage and the relative 3dB advantage of heterodyne over division in the sample-power-constrained case remains the same as the local SNR case, since at this limit the conversion factor from local SNR to global SNR is identical for both the heterodyne and the division receivers.

\hw{
\subsection{SNR of general sensing tasks}
\label{sec:SNR_phase_supp}
We have defined the SNR for estimating the absorption $\sqrt{\kappa}$. It is easy to generalize the SNR for general sensing tasks of estimating an arbitrary unknown parameter $\lambda$, since the only difference lies in the dependence of the readout mean value $\mu$ on $\lambda$. For local SNR, one can connect the general $\lambda$-estimation SNR (${\rm SNR}^2_{\lambda}$) to the $\sqrt\kappa$-estimation SNR (${\rm SNR}^2_{\sqrt{\kappa}}$) using the following formulas:
\be 
{\rm SNR}^2_{\sqrt{\kappa}}=\frac{(\partial_{\sqrt{\kappa}} \mu)^2}{\sigma^2} = \frac{(\partial_{\sqrt{\kappa}} \mu)^2}{(\partial_{\lambda} \mu)^2}\frac{(\partial_{\lambda} \mu)^2}{\sigma^2}
=\frac{(\partial_{\sqrt{\kappa}} \mu)^2}{(\partial_{\lambda} \mu)^2}{\rm SNR}^2_{\lambda}\,.
\ee
And similarly for global SNR
\be 
{\rm SNR}^2_{\sqrt{\kappa}}=\frac{(\mu|_{\kappa}-\mu|_{\kappa=\kappa_0})^2}{(\mu|_{\lambda}-\mu|_{\lambda=\lambda_0})^2}{\rm SNR}^2_{\lambda}.
\ee
where $\kappa_0, \lambda_0$ are the reference constants. 
}

\hw{
Besides absorption sensing, a typical task is phase sensing, where $\lambda=\theta$. For simplicity, we take the local SNR of heterodyne detection as an example to elaborate the connection in SNR.  
We estimate the phase at the $m$th line, i.e. $\theta=\theta_m$, Eq.~\eqref{d_m_het_app} gives the readout mean value
\be 
\mu=\expval{\hat{d}_m}=\sqrt{\kappa_m}e^{i\theta} A_m B_m^\star+\sqrt{\kappa_{-m}}e^{-i\theta_{-m}} A_{-m}^\star B_{-m}.
\ee 
Then $(\partial_{\theta} \mu)^2=\kappa_m |A_mB_m|^2$. On the other hand, considering estimation of the $m$th line absorption $\kappa=\kappa_m$, we have $(\partial_{\sqrt{\kappa}} \mu)^2= |A_mB_m|^2$. Thus the phase-sensing SNR can be obtained from the absorption-sensing SNR via
\be 
{\rm SNR}^2_{\theta}=\kappa \cdot {\rm SNR}^2_{\sqrt{\kappa}}\,.
\label{eq:SNR_phase_supp}
\ee
This recovers the results derived in Ref.~\cite{shi2023entanglement}.
}

\section{Combine then pass through sample}
\label{sec:combine_pass}

As shown in Fig.~\ref{fig:schematic}(b) of the main text, two input combs $\hat A(t)$ and $\hat B(t)$ first interfere at a balanced beamsplitter, leading to outputs
\bal 
\hat A'(t)&=\frac{\hat A(t)+\hat B(t)}{\sqrt{2}}\,,\\
\hat B'(t)&=\frac{\hat A(t)-\hat B(t)}{\sqrt{2}}\,.
\eal 
Then one of the outputs $\hat A'(t)$ probes the sample, which incurs the input-output relation in Eq.~\eqref{loss_channel} of the main text to produce $\calL[\hat A'(t)]$.
The power on the sample can be obtained as
$({P_A+P_B})/{2}$,
where $P_A,P_B$ are given in Eqs.~\eqref{power_AB}. At the photon detectors, the time domain photocurrents can be expressed by the operators
\bal 
\hat I_A(t)
&=\calL[\hat A^{\prime}(t)]^\dagger\calL[\hat A'(t)]\\
&=\frac{1}{2}[\calL(\hat A(t))+\calL(\hat B(t))]^\dagger \calL[\hat A(t)+\hat B(t)],
\label{IAt}
\eal
\bal 
\hat I_B(t)&=\hat B^{\prime\dagger}(t)\hat B'(t)\\
&=\frac{1}{2}[\hat A^\dagger(t)-\hat B^\dagger(t)][\hat A(t)-\hat B(t)].
\label{IBt}
\eal
In the data processing, one performs Fourier transform of the photocurrent  and applies a bandpass filter that only keeps the intermediate-frequency components at $\sim\Delta \omega_r$, filtering out zero-frequency DC components and high-frequency components at $\sim\omega_r$.

To begin with, we solve the modulation in the mean photocurrent. Eq.~\eqref{IAt} and Eq.~\eqref{IBt} along with Eq.~\eqref{eq_combAB_mean} and Eq.~\eqref{eq_combAB_mean_loss} yield
\ba 
&\expval{\hat I_A(t)}&=\frac{1}{2T}\sum_{n=-N}^N e^{in \Delta \omega_rt} \kappa_n A_n^\star B_n+c.c.+\cdots
\\
&\expval{\hat I_B(t)}&=-\frac{1}{2T}\sum_{n=-N}^N e^{in \Delta \omega_rt} A_n^\star B_n+c.c.+\cdots
\ea 
where `$\cdots$' denote DC term and high-frequency terms. After the bandpass filter, in the frequency domain, we obtain the intermediate-frequency components as
\be
\hat I_X(m\Delta \omega_r)\equiv \int e^{im \Delta\omega_r t} \hat I_X(t) dt, 
\ee 
for $X=A,B$.
Their mean values are
\begin{subequations}
\ba 
&\!\!\!\!\!\!\expval{\hat I_A(m\Delta \omega_r)}=&\frac{1}{2}\left(\kappa_m A_m B_m^\star+\kappa_{-m} A_{-m}^\star B_{-m}\right)
\\
&\!\!\!\!\!\!\expval{\hat I_B(m\Delta \omega_r)}=&\frac{1}{2}\left(-A_mB_m^\star-A_{-m}^\star B_{-m}\right) .
\ea 
\label{eq:I_mf_mean_supp}
\end{subequations} 

Now we consider the noise. We can decompose the photocurrent into the mean and a zero-mean additive noise $\Delta \hat{I}_X$ as 
\be 
\hat{I}_X(t)=\expval{\hat{I}_X(t)}+\Delta \hat{I}_X(t),
\ee 
and the same for the Fourier spectrum in frequency domain 
\be 
\hat I_X(m\Delta \omega_r)=\expval{\hat I_X(m\Delta \omega_r)}+\Delta \hat I_X(m\Delta \omega_r).
\label{I_X_approximate}
\ee 
In the limit of $A_n,B_n\gg1$,
we have
\begin{align}
&\Delta \hat I_B(t)=\frac{1}{2}\expval{\hat A^\dagger(t)-\hat B^\dagger(t)}
\nonumber
\\
&\times \frac{1}{\sqrt{T}}\sum_{n=-N}^N\sum_{m=-2N}^{2N} e^{-i (n\omega_r+m\Delta\omega_r) t}(\hat \calA_{n,m}-\hat \calB_{n,m})+ c.c.
\\
&=\frac{1}{2}[\frac{1}{\sqrt{T}}\sum_{n=-N}^N e^{i n(\omega_r+\Delta\omega_r) t} A_n^\star-\frac{1}{\sqrt{T}}\sum_{n=-N}^N e^{i n\omega_r t} B_n^\star]
\nonumber
\\
&\times \frac{1}{\sqrt{T}}\sum_{n=-N}^N\sum_{m=-2N}^{2N} e^{-i (n\omega_r+m\Delta\omega_r) t}(\hat \calA_{n,m}-\hat \calB_{n,m})+ c.c.
\end{align}  
where we used Eq.~\eqref{eq_combAB_mean} in the second step.

In frequency domain, 
\begin{align}
&\Delta \hat I_B(m \Delta \omega_r)=
\nonumber
\\
&\frac{1}{2}\sum_{n=-N}^N \left[A_n^\star (\hat \calA_{n,n+m}-\hat \calB_{n,n+m})-B_n^\star(\hat \calA_{n,m}-\hat \calB_{n,m}) \right]
\nonumber
\\
&+\left[A_n (\hat \calA_{n,n-m}^\dagger-\hat \calB_{n,n-m}^\dagger)-B_n(\hat \calA_{n,-m}^\dagger-\hat \calB_{n,-m}^\dagger) \right]\,,
\label{eq:photocurrent_spectrum_B_supp}
\end{align}
similarly
{
\small
\begin{align}
&\Delta \hat I_A(m \Delta \omega_r)=
\nonumber
\\
&=\frac{1}{2}\sum_{n=-N}^N \!\kappa_n \! \left[A_n^\star (\hat \calA_{n,n+m}+\hat \calB_{n,n+m}) \!+\! B_n^\star(\hat \calA_{n,m}+\hat \calB_{n,m}) \right]\nonumber
\\
&+\frac{1}{\sqrt{2}}\sum_{n=-N}^N \sqrt{\kappa_n(1-\kappa_n)} \left[A_n^\star \hat \calV_{n,n+m}+B_n^\star \hat \calV_{n,m}\right]
\nonumber
\\
&+\frac{1}{2}\!\sum_{n=-N}^N \!\kappa_n \!\left[\! A_n (\hat \calA_{n,n-m}^\dagger\!+\!\hat \calB_{n,n-m}^\dagger)\! +\! B_n(\hat \calA_{n,-m}^\dagger\!+\!\hat \calB_{n,-m}^\dagger) \right]
\nonumber
\\
&+\frac{1}{\sqrt{2}}\sum_{n=-N}^N \sqrt{\kappa_n(1-\kappa_n)} \left[A_n \hat \calV_{n,n-m}^\dagger+B_n \hat \calV_{n,-m}^\dagger\right]\,.
\label{eq:photocurrent_spectrum_A_supp}
\end{align}
}
Here $\hat \calV_{n,m}$ are thermal environment modes at frequency $n\omega_r+m\Delta \omega_r$, which is in thermal state of mean photon number 
\be 
E_{n}=\frac{1}{e^{\hbar n \omega_r/k_BT}-1}\,,
\ee 
subject to the Bose-Einstein distribution.


\subsection{Division data processing}
\label{supp:division}
From Eqs.~\eqref{eq:I_mf_mean_supp}, we can extract information about the absorption $\kappa_m$ and $\kappa_{-m}$ by measuring the ratio of the mean photocurrent spectra
\ba 
-\frac{\expval{\hat I_A(m\Delta \omega_r)}}{\expval{\hat I_B(m\Delta \omega_r)}}\equiv r_m=c_{+,m}\kappa_m+c_{-,m}\kappa_{-m}
\label{eq:rm_app}
\ea 
where $c_{+,m}=A_{m} B_{m}^\star/(A_mB_m^\star+A_{-m}^\star B_{-m})$ and $c_{-,m}=A_{-m}^\star B_{-m}/(A_mB_m^\star+A_{-m}^\star B_{-m})$ are $O(1)$ parameters. For example, for symmetric comb $A_n=A_{-n}^*,B_n=B_{-n}^*$, we have
$r_m=(\kappa_m+\kappa_{-m})/2$. Here we observe that the comb spectrum is automatically calibrated: an unknown comb spectrum does not affect the measurement result. \hw{To extract information from both $\kappa_m$ and $\kappa_{-m}$, various strategies can be adopted to eliminate the degeneracy of information from $\pm m$ frequency components, which we will address later in \ref{sec:aliase_supp}.} 

In general, the ratio including noise is defined by the quantum operator
\be 
\hat r_m\equiv -\frac{\hat I_A(m\Delta \omega_r)}{\hat I_B(m\Delta \omega_r)}.
\label{eq:div_rm_supp}
\ee 
\hw{Here $1/\hat I_B(m\Delta \omega_r)$ represents the Moore–Penrose pseudoinverse of operator $\hat I_B(m\Delta \omega_r)$.}
Note that in general $\expval{\hat r_n}\neq -\expval{\hat I_A(m\Delta \omega_r)}/\expval{\hat I_B(m\Delta \omega_r)}$, because the mean value of the ratio is not necessarily equal to the ratio of mean values. In the strong comb line limit $A_n,B_n\gg1$,
\begin{align}
\hat r_m&=-\frac{\expval{\hat I_A(m\Delta \omega_r)}+\Delta \hat I_A(m\Delta \omega_r)}{\expval{\hat I_B(m\Delta \omega_r)}+\Delta \hat I_B(m\Delta \omega_r)}
\\
&\simeq r_m+\frac{\Delta \hat I_A(m\Delta \omega_r)+r_m\Delta \hat I_B(m\Delta \omega_r)}{(A_mB_m^\star+A_{-m}^\star B_{-m})/2 }.
\end{align}
To the leading order, the mean is
\bal 
\expval{\hat r_m}&\simeq {r_m} = \frac{\kappa_m A_m B_m^\star+\kappa_{-m} A_{-m}^\star B_{-m}}{A_mB_m^\star+A_{-m}^\star B_{-m}},
\eal
and the variance is
\begin{align}
{\rm var}(\hat r_m)=\frac{{\rm var}[\Delta \hat I_A(m\Delta \omega_r)+r_m\Delta \hat I_B(m\Delta \omega_r)]}{|A_mB_m^\star+A_{-m}^\star B_{-m}|^2/4}.
\label{eq:rm_var_app}
\end{align}
In the scenario of a single absorption line considered in the main text, the SNR for estimation of $\sqrt{\kappa_m}$ is given by $\kappa_m/{\rm var}(\hat r_m)$, as the mean $\expval{\hat r_m}=(\kappa_m+1)/2$ and the chain rule in Eq.~\eqref{eq:SNR_point} of the main text. On the other hand, the global SNR, defined in Eq.~\eqref{eq:SNR_global} of the main text, is $(\kappa_m-1)^2/(4{\rm var}(\hat r_m))$.

The full formula of the variance is lengthy, in the main text we focus on the lossless limit to first obtain some intuition, where $r_m\simeq 1$ as $\kappa_m \simeq \kappa_{-m}\simeq 1$. In this case,
\begin{align}
{\rm var}(\hat r_m)\simeq\frac{{\rm var}[\Delta \hat I_A(m\Delta \omega_r)+\Delta \hat I_B(m\Delta \omega_r)]}{|A_mB_m^\star+A_{-m}^\star B_{-m}|^2/4}.
\end{align}
At the lossless limit, from Eqs.~\eqref{eq:photocurrent_spectrum_B_supp}~\eqref{eq:photocurrent_spectrum_A_supp}, we can derive the numerator of the variance Eq.~\eqref{eq:rm_var_app} as
\begin{align}
&\Delta \hat I_A(m \Delta \omega_r)+\Delta \hat I_B(m \Delta \omega_r)=
\nonumber
\\
&\sum_{n=-N}^N \left[A_n^\star \hat \calA_{n,n+m}+A_n \hat \calA_{n,n-m}^\dagger+B_n^\star\hat \calB_{n,m}+B_n\hat \calB_{n,-m}^\dagger  \right]
\end{align}

For general lossy case, we derive
\begin{widetext}
\begin{align} 
\Delta \hat I_A(m\Delta \omega_r)+{r_m}\Delta \hat I_B(m\Delta \omega_r)=
&\frac{1}{2}\sum_{n=-N}^N (\kappa_n+{r_m})
\left[
A_n^\star \hat \calA_{n,n+m}+
A_n \hat \calA_{n,n-m}^\dagger
+B_n^\star\hat \calB_{n,m}
+B_n\hat \calB_{n,-m}^\dagger
\right]
\nonumber
\\
&+\frac{1}{2}\sum_{n=-N}^N 
(\kappa_n-{r_m})
\left[
A_n^\star \hat \calB_{n,n+m}
+A_n \hat \calB_{n,n-m}^\dagger
+B_n^\star\hat \calA_{n,m} 
+B_n\hat \calA_{n,-m}^\dagger
\right]
\nonumber
\\
&+\frac{1}{\sqrt{2}}\sum_{n=-N}^N \sqrt{\kappa_n(1-\kappa_n)} \left[
A_n^\star \hat \calV_{n,n+m}
+A_n \hat \calV_{n,n-m}^\dagger
+B_n^\star \hat \calV_{n,m}
+B_n \hat \calV_{n,-m}^\dagger\right]
\label{IA_IB_division_sum}
\end{align}
\end{widetext}

\subsubsection{Intra-comb-line squeezing}
\label{app:divRx_intrasqz}
Now we suppress the readout variance, by sideband two-mode squeezings between the sideband noise modes for each line individually. Assuming perfect phase locking that $A_n, B_n$ are real and loss is weak $\kappa_n\to 1$, the division information processing keeps only the self-beating noise terms (where the noises beat with the comb lines at the same comb), and the cross-beating noise terms, i.e. the second row in Eq.~\eqref{IA_IB_division_sum}, vanish. In this case we can suppress the self-beating noises by two-mode squeezing between $(\hat \calA_{n,n+m}, \hat \calA_{n,n-m})$ and two-mode squeezing between $(\hat \calB_{n,m}, \hat \calB_{n,-m})$ for all relevant $n,m$. From the squeezed quadrature variances in Eqs.~\eqref{sqz_var_intra_supp}\eqref{antisqz_var_intra_supp}, we have the squeezed variances
\bal 
{\rm var} (\hat\calA_{n,n+m} \!+\! \hat\calA_{n,n-m}^\dagger)&= 1/G_{A,n}\,,\\
{\rm var} (  \hat\calB_{n,m} \!+\!\hat\calB_{n,-m}^\dagger )&= 1/G_{B,n}.
\eal 

However, when loss is significant, $\kappa_n<1$, the cross-beating noises are not negligible. For the two-mode squeezing above, the resulting two-mode squeezing is mismatched for the cross-beating noises paired in the form of $(\hat\calA_{n,-m}, \hat\calA_{n,m}) $ and $ (\hat \calB_{n,n+m}, \hat \calB_{n,n-m})$. The two-mode squeezing mismatching invokes amplified spontaneous emission noises, as a result of tracing out the environment in a parametric amplifier that leads to a bosonic amplifier channel with quantum amplification noise~\cite{weedbrook2012gaussian}:
\small
\bal 
{\rm var} ( \hat\calA_{n,-m}^\dagger\!+\!\hat\calA_{n,m} )&\!\to\! G_{A,n}^\prime\equiv 1\!+\!2N^{\rm amp}_{A,n}\!=\!\frac{1}{2}(G_{A,n}\!+\!\frac{1}{G_{A,n}})\\
{\rm var} (\hat \calB_{n,n+m}+ \hat \calB_{n,n-m}^\dagger)&\!\to\! G_{B,n}^\prime\equiv 1\!+\!2N^{\rm amp}_{B,n}\!=\!\frac{1}{2}(G_{B,n}\!+\!\frac{1}{G_{B,n}})
\eal 
\normalsize
where $N^{\rm amp}_{X,n}=\frac{G_{X,n}+\frac{1}{G_{X,n}}-2}{4}$, $X=A,B$. 


Given the two-mode squeezing above, the numerator of the variance Eq.~\eqref{eq:rm_var_app} is
\begin{align}
&{\rm var}[\Delta \hat I_A(m\Delta \omega_r)+{r_m}\Delta \hat I_B(m\Delta \omega_r)]=
\nonumber
\\
&\frac{1}{4}\sum_{n=-N}^N \Bigg\{
(\kappa_n+{r_m})^2 \left(\frac{|A_n|^2}{G_{A,n}}+\frac{|B_n|^2}{G_{B,n}}\right)
\nonumber
\\
&
+(\kappa_n-{r_m})^2(|A_n|^2 G_{B,n}^\prime+|B_n|^2 G_{A,n}^\prime)
\nonumber
\\
&+2\kappa_n(1-\kappa_n)(|A_n|^2+|B_n|^2)(1+2E_n)\Bigg\}
\end{align}
The final formula of the estimator variance is
\begin{widetext}
\small
\begin{align}
{\rm var}(\hat r_m)&=\frac{\sum_{n=-N}^N\!\left(\kappa_n+{r_m}\right)^2 \left(\frac{|A_n|^2}{G_{A,n}}+\frac{|B_n|^2}{G_{B,n}}\right)
\!+\!\left(\kappa_n-{r_m}\right)^2\left(|A_n|^2 G_{B,n}^\prime+|B_n|^2 G_{A,n}^\prime\right)
\!+\!2\kappa_n\left(1\!-\!\kappa_n\right)\left(|A_n|^2\!+\! |B_n|^2\right)\left(1\!+\! 2E_n\right)}{|A_mB_m^\star+A_{-m}^\star B_{-m}|^2}.
\label{eq:var_div_intra_supp}
\end{align}
\normalsize
\end{widetext}
Here we observe that the protocol is resistant against phase noise in $\theta_n$, while fragile to non-uniform absorption spectrum $\kappa_n$: the amplified spontaneous noises $\propto G_{X,n}'$ mix in from all lines, except for a uniform absorption spectrum $\kappa_n=r_m$ for all $n$. We use this formula to produce the numerical evaluations of SNR in the maintext.

The loss-sensitiveness of division receiver makes it incompatible with single-side combs, which have $A_{\ell}B_{\ell}^\star=0$ for $\ell<0$. With such single-side combs, we have $r_m=\kappa_m$ as $c_{+,m}=1, c_{-,m}=0$. The variance contains the $\kappa_n-\kappa_m$ terms which are equal to $\kappa_n$ for $m<0$, significantly non-zero. The amplified noises due to the two-mode squeezing mismatching then always mix in.

\hw{Now consider uniform squeezing and comb line spectrum $G_{A,n}=G_A,G_{B,n}=G_B$, $A_n=A, B_n=B$ and the single-absorption-line case, i.e. $\kappa_m=\kappa, \kappa_{n\neq m}=1$. Then ${r_m} = (\kappa+1)/2$, the division ratio variance Eq.~\eqref{eq:var_div_intra_supp} becomes
\begin{widetext}
\bal
{\rm var}(\hat r_m)&=
\frac{1}{4} \Bigg\{ (M-1) \left[\frac{1}{4} (1-\kappa )^2 \left(\frac{G_A+\frac{1}{G_A}}{2 A^2}+\frac{G_B+\frac{1}{G_B}}{2 B^2}\right)+\frac{1}{4} (\kappa +3)^2
   \left(\frac{1}{A^2 G_B}+\frac{1}{B^2 G_A}\right)\right]\\
  &\quad\quad  +\left(\kappa -\frac{\kappa +1}{2} \right)^2 \left(\frac{G_A+\frac{1}{G_A}}{2
   A^2}+\frac{G_B+\frac{1}{G_B}}{2 B^2}\right)+\left(\kappa +\frac{\kappa +1}{2}\right)^2 \left(\frac{1}{A^2 G_B}+\frac{1}{B^2 G_A}\right) \\
   &\quad\quad  +2  \kappa  (1-\kappa )
   \left(2 E_m+1\right) \left(\frac{1}{A^2}+\frac{1}{B^2}\right)\Bigg\}
   \label{eq:var_div_intra_singleline_supp}
\eal
\end{widetext}
for the division readout at the absorption line.
Take the limit $M\gg G$ and note that the Fisher-type local SNR is defined by Eq.~\eqref{eq:SNR_point} with the signal $ 
\expval{\hat r_m}\simeq {r_m}$ and $|\partial_{\sqrt{\kappa}} \expval{\hat r_m}|^2=\kappa$, we recover Eq.~\eqref{SNR_div_single_line} in the maintext.
}

\hw{We can generalize the above single-absorption-line case to the $\gamma M$-uniform-absorption-line case with uniform transmissivity $\kappa$. We assume $\gamma$ is small and the $\gamma M$ absorption lines locates at the same side of the carrier, then ${r_m} = (\kappa+1)/2$ for all the $\gamma M$ absorption lines. Now the variance Eq.~\eqref{eq:var_div_intra_supp} becomes
\begin{widetext}
\bal
{\rm var}(\hat r_m)&=
\frac{1}{4} \Bigg\{ (1-\gamma)M \left[\frac{1}{4} (1-\kappa )^2 \left(\frac{G_A+\frac{1}{G_A}}{2 A^2}+\frac{G_B+\frac{1}{G_B}}{2 B^2}\right)+\frac{1}{4} (\kappa +3)^2
   \left(\frac{1}{A^2 G_B}+\frac{1}{B^2 G_A}\right)\right]\\
  &\quad\quad  +\gamma M\left[\left(\kappa -\frac{\kappa +1}{2} \right)^2 \left(\frac{G_A+\frac{1}{G_A}}{2
   A^2}+\frac{G_B+\frac{1}{G_B}}{2 B^2}\right)+\left(\kappa +\frac{\kappa +1}{2}\right)^2 \left(\frac{1}{A^2 G_B}+\frac{1}{B^2 G_A}\right) \right]\\
   &\quad\quad  +\gamma M\left[2  \kappa  (1-\kappa )
   \left(2 E_m+1\right) \left(\frac{1}{A^2}+\frac{1}{B^2}\right)\right]\Bigg\}
\label{eq:SNR_div_intra_gammaM_supp}
\eal
\end{widetext}
for the division readouts at all the $\gamma M$ absorption lines. 
}

\subsubsection{Cross-comb-line entanglement}
Alternatively, we can implement noise reduction by a broadband two-mode squeezing between the noise modes paired around the carrier frequency across the whole comb, which we denote as cross-comb-line entanglement. In the lossless limit, this can be easily seen by re-pairing the squeezed modes around the carrier frequency 
\begin{align}
&\Delta \hat I_A(m \Delta \omega_r)+\Delta \hat I_B(m\Delta \omega_r)
\nonumber
\\
&=\frac{1}{T}\sum_{n=1}^N \left[A_n^\star \hat \calA_{n,n+m}+A_{n} \hat \calA_{n,n-m}^\dagger+B_n^\star\hat \calB_{n,m}+B_{n}\hat \calB_{n,-m}^\dagger  \right]
\nonumber\\
&\quad +\frac{1}{T}\sum_{n=1}^{N} \Big[A_{-n}^\star \hat \calA_{-n,-n+m}+A_{-n} \hat \calA_{-n,-n-m}^\dagger
\nonumber\\
&\quad +B_{-n}^\star\hat \calB_{-n,m}+B_{-n}\hat \calB_{-n,-m}^\dagger  \Big]
\nonumber\\
&=\frac{1}{T}\sum_{n=1}^N \Big[(A_n^\star \hat \calA_{n,n+m}+A_{-n} \hat \calA_{-n,-n-m}^\dagger)
\nonumber\\
&\qquad +(A_n \hat \calA_{n,n-m}^\dagger+A_{-n}^\star \hat \calA_{-n,-n+m})
\nonumber\\
&\qquad +(B_n^\star\hat \calB_{n,m}+B_{-n}\hat \calB_{-n,-m}^\dagger ) 
\nonumber\\
&\qquad + (B_n\hat \calB_{n,-m}^\dagger +B_{-n}^\star\hat \calB_{-n,m}) \Big]
\nonumber\\
\end{align}
If $A_n=A_{-n}^*, B_n=B_{-n}^*$, then a single pump line at the carrier frequency can squeeze $(\hat \calA_{n,n+m}+\hat \calA_{-n,-n-m}^\dagger)$, $(\hat \calB_{n,m}+\hat \calB_{-n,-m}^\dagger )$, etc., to reduce the readout noise. Such two-mode squeezing centered at the carrier forms a broadband entanglement across the whole comb.

To describe the noises of the entangled pairs, we define quadrature operators $\hat q^\calA_{n,m}\equiv \sqrt{2}\Re \hat \calA_{n,m}$, $\hat p^\calA_{n,m}\equiv \sqrt{2}\Im \hat \calA_{n,m}$ for $\hat \calA$, and the `$+$' common mode and `$-$' differential mode, with the quadrature operators $\hat q^{\calA\pm}_{n,m}\equiv \hat q^\calA_{n,m}\pm\hat q^\calA_{-n,-m}, \hat p^{\calA\pm}_{n,m}\equiv \hat p^\calA_{n,m}\pm\hat p^\calA_{-n,-m}$. Then two-mode squeezing can be described by the squeezed and anti-squeezed quadratures
\bal 
{\rm var}\{\hat q_{n,m}^{\calA+}\}={\rm var}\{\hat p_{n,m}^{\calA-}\}&=1/G_{A,n}\\
{\rm var}\{\hat q_{n,m}^{\calA-}\}=
{\rm var}\{\hat p_{n,m}^{\calA+}\}&=G_{A,n}\,.
\label{eq:EPR_quadrature_noise}
\eal 
Similar for $\hat \calB$. 

In the general lossy case, the noise modes in Eq.~\eqref{IA_IB_division_sum} can be paired as
\begin{widetext}
\begin{align} 
&\Delta \hat I_A(m\Delta \omega_r)+{r_m}\Delta \hat I_B(m\Delta \omega_r)=
\\
&=\frac{1}{2}\sum_{n=-N}^N 
\left[
(\kappa_n+{r_m}) A_n^\star \hat \calA_{n,n+m}
+(\kappa_{-n}+{r_m}) A_{-n} \hat \calA_{-n,-n-m}^\dagger
+(\kappa_n+{r_m}) B_n^\star\hat \calB_{n,m}
+(\kappa_{-n}+{r_m}) B_{-n}\hat \calB_{-n,-m}^\dagger
\right]
\nonumber
\\
&\quad  +\frac{1}{2}\sum_{n=-N}^N 
\left[
(\kappa_n-{r_m})A_n^\star \hat \calB_{n,n+m}
+(\kappa_{-n}-{r_m})A_{-n} \hat \calB_{-n,-n-m}^\dagger
+(\kappa_n-{r_m})B_n^\star\hat \calA_{n,m} 
+(\kappa_{-n}-{r_m})B_{-n}\hat \calA_{-n,-m}^\dagger
\right]
\nonumber
\\
&\quad  +\frac{1}{\sqrt{2}}\sum_{n=-N}^N \sqrt{\kappa_n(1-\kappa_n)} \left[
A_n^\star \hat \calV_{n,n+m}
+A_n \hat \calV_{n,n-m}^\dagger
+B_n^\star \hat \calV_{n,m}
+B_n \hat \calV_{n,-m}^\dagger\right].
\label{IA_IB_division_sum_cross}
\end{align}
It can be decomposed into two independent parts via $\Delta \hat I_A(m\Delta \omega_r)+{r_m}\Delta \hat I_B(m\Delta \omega_r)=\hat\Sigma_Q(m\Delta \omega_r)+i\hat\Sigma_P(m\Delta \omega_r)$: $\hat\Sigma_Q$ contributed by position quadratures $\{\hat q_{n,m}\}$, and $\hat\Sigma_P$ contributed by momentum quadratures $\{\hat p_{n,m}\}$, defined as
\begin{align} 
&\sqrt{2}\hat\Sigma_Q(m\Delta \omega_r)
\\
&=\frac{1}{2}\sum_{n=-N}^N 
\left[
(\kappa_n+{r_m}) A_n^\star \hat q^\calA_{n,n+m}
+(\kappa_{-n}+{r_m}) A_{-n} \hat q^\calA_{-n,-n-m}
+(\kappa_n+{r_m}) B_n^\star\hat q^\calB_{n,m}
+(\kappa_{-n}+{r_m}) B_{-n}\hat q^\calB_{-n,-m}
\right]
\nonumber
\\
&\quad  +\frac{1}{2}\sum_{n=-N}^N 
\left[
(\kappa_n-{r_m})A_n^\star \hat q^\calB_{n,n+m}
+(\kappa_{-n}-{r_m})A_{-n} \hat q^\calB_{-n,-n-m}
+(\kappa_n-{r_m})B_n^\star\hat q^\calA_{n,m} 
+(\kappa_{-n}-{r_m})B_{-n}\hat q^\calA_{-n,-m}
\right]
\nonumber
\\
&\quad  +\frac{1}{\sqrt{2}}\sum_{n=-N}^N \sqrt{\kappa_n(1-\kappa_n)} \left[
A_n^\star \hat \calV_{n,n+m}
+A_n \hat \calV_{n,n-m}^\dagger
+B_n^\star \hat \calV_{n,m}
+B_n \hat \calV_{n,-m}^\dagger\right]\,,
\end{align}
\begin{align} 
&\sqrt{2}\hat\Sigma_P(m\Delta \omega_r)
\\
&=\frac{1}{2}\sum_{n=-N}^N 
\left[
(\kappa_n+{r_m}) A_n^\star \hat p^\calA_{n,n+m}
-(\kappa_{-n}+{r_m}) A_{-n} \hat p^\calA_{-n,-n-m}
+(\kappa_n+{r_m}) B_n^\star\hat p^\calB_{n,m}
-(\kappa_{-n}+{r_m}) B_{-n}\hat p^\calB_{-n,-m}
\right]
\nonumber
\\
&\quad  +\frac{1}{2}\sum_{n=-N}^N 
\left[
(\kappa_n-{r_m})A_n^\star \hat p^\calB_{n,n+m}
-(\kappa_{-n}-{r_m})A_{-n} \hat p^\calB_{-n,-n-m}
+(\kappa_n-{r_m})B_n^\star\hat p^\calA_{n,m} 
-(\kappa_{-n}-{r_m})B_{-n}\hat p^\calA_{-n,-m}
\right]
\nonumber
\\
&\quad  +\frac{1}{\sqrt{2}}\sum_{n=-N}^N \sqrt{\kappa_n(1-\kappa_n)} \left[
A_n^\star \hat \calV_{n,n+m}
+A_n \hat \calV_{n,n-m}^\dagger
+B_n^\star \hat \calV_{n,m}
+B_n \hat \calV_{n,-m}^\dagger\right].
\end{align}
Thus ${\rm var}\{\Delta \hat I_A(m\Delta \omega_r)+{r_m}\Delta \hat I_B(m\Delta \omega_r)\}={\rm var}\hat\Sigma_Q(m\Delta \omega_r)+ {\rm var}\hat\Sigma_P(m\Delta \omega_r)$.
The variances of the two parts are
\bal
~&{\rm var}\hat\Sigma_Q(m\Delta \omega_r)={\rm var}\hat\Sigma_P(m\Delta \omega_r)
\\
&=\frac{1}{2}{\rm var}\Bigg\{
\\
&\quad \frac{1}{2}\sum_{n=-N}^N 
\Big[
(\kappa_n+{r_m})A_n^\star  \frac{\hat q^{\calA+}_{n,n+m}+\hat q^{\calA-}_{n,n+m}}{2}
+(\kappa_{-n}+{r_m}) A_{-n} \frac{\hat q^{\calA+}_{n,n+m}-\hat q^{\calA-}_{n,n+m}}{2}\\
&\qquad \qquad 
+(\kappa_n+{r_m}) B_n^\star \frac{\hat q^{\calB+}_{n,m}+\hat q^{\calB-}_{n,m}}{2}
+(\kappa_{-n}+{r_m}) B_{-n} \frac{\hat q^{\calB+}_{n,m}-\hat q^{\calB-}_{n,m}}{2}
\Big]
\\
&\quad  +\frac{1}{2}\sum_{n=-N}^N 
\Big[
(\kappa_n-{r_m})A_n^\star \frac{\hat q^{\calB+}_{n,n+m}+\hat q^{\calB-}_{n,n+m}}{2}
+(\kappa_{-n}-{r_m})A_{-n} \frac{\hat q^{\calB+}_{n,n+m}-\hat q^{\calB-}_{n,n+m}}{2}\\
&\qquad \qquad
+(\kappa_n-{r_m})B_n^\star \frac{\hat q^{\calA+}_{n,m}+\hat q^{\calA-}_{n,m}}{2}
+(\kappa_{-n}-{r_m})B_{-n}\frac{\hat q^{\calA+}_{n,m}-\hat q^{\calA-}_{n,m}}{2}
\Big]
\\
&\quad  +\frac{1}{\sqrt{2}}\sum_{n=-N}^N \sqrt{\kappa_n(1-\kappa_n)} \left[
A_n^\star \hat \calV_{n,n+m}
+A_n \hat \calV_{n,n-m}^\dagger
+B_n^\star \hat \calV_{n,m}
+B_n \hat \calV_{n,-m}^\dagger\right]
\Bigg\}
\\
&=\frac{1}{8}\sum_{n=-N}^N 
\Big[
  \left|\frac{(\kappa_n+{r_m})A_n^\star+(\kappa_{-n}+{r_m}) A_{-n}}{2}\right|^2\frac{1}{G_{A,n}}
+ \left|\frac{(\kappa_n+{r_m})A_n^\star-(\kappa_{-n}+{r_m}) A_{-n}}{2}\right|^2G_{A,n} \\
&\qquad \qquad \quad
+ \left|\frac{(\kappa_n+{r_m}) B_n^\star+(\kappa_{-n}+{r_m}) B_{-n} }{2}\right|^2\frac{1}{G_{B,n}}
+ \left|\frac{(\kappa_n+{r_m}) B_n^\star-(\kappa_{-n}+{r_m}) B_{-n} }{2}\right|^2G_{B,n}
\\
&\qquad \qquad \quad
+ \left|\frac{(\kappa_n-{r_m})A_n^\star+(\kappa_{-n}-{r_m}) A_{-n}}{2}\right|^2\frac{1}{G_{B,n}}
+ \left|\frac{(\kappa_n-{r_m})A_n^\star-(\kappa_{-n}-{r_m}) A_{-n}}{2}\right|^2G_{B,n} \\
&\qquad \qquad \quad
+ \left|\frac{(\kappa_n-{r_m}) B_n^\star+(\kappa_{-n}-{r_m}) B_{-n} }{2}\right|^2\frac{1}{G_{A,n}}
+ \left|\frac{(\kappa_n-{r_m}) B_n^\star-(\kappa_{-n}-{r_m}) B_{-n} }{2}\right|^2G_{A,n}
\\
&\qquad \qquad \quad  
+2\kappa_n(1-\kappa_n)(|A_n|^2+|B_n|^2)(1+2E_n) \Big]\,.
\label{eq:CrossEntangle_var_QP_app}
\eal

Furthermore, if the comb envelop is real $A_n=A_{-n}^\star$, $B_n=B_{-n}^\star$, then the formulas can be further simplified:
\bal
~&{\rm var}\hat\Sigma_Q(m\Delta \omega_r)|_{A_n=A_{-n}^\star, B_n=B_{-n}^\star}\\
&={\rm var}\hat\Sigma_P(m\Delta \omega_r)|_{A_n=A_{-n}^\star, B_n=B_{-n}^\star}\\
&=\frac{1}{8}\sum_{n=-N}^N 
\Big[
  \left|\frac{\kappa_n+\kappa_{-n}}{2}+{r_m}\right|^2\frac{|A_n|^2}{G_{A,n}}
+ \left|\frac{\kappa_n-\kappa_{-n} }{2}\right|^2|A_n|^2G_{A,n} \\
&\qquad \qquad \quad
+ \left|\frac{\kappa_n+\kappa_{-n} }{2}+{r_m}\right|^2\frac{|B_n|^2}{G_{B,n}}
+ \left|\frac{\kappa_n-\kappa_{-n}  }{2}\right|^2|B_n|^2G_{B,n}
\\
&\qquad \qquad \quad
+ \left|\frac{\kappa_n+\kappa_{-n} }{2}-{r_m}\right|^2\frac{|A_n|^2}{G_{B,n}}
+ \left|\frac{\kappa_n-\kappa_{-n} }{2}\right|^2|A_n|^2G_{B,n} \\
&\qquad \qquad \quad
+ \left|\frac{\kappa_n+\kappa_{-n}  }{2}-{r_m}\right|^2\frac{|B_n|^2}{G_{A,n}}
+ \left|\frac{\kappa_n -\kappa_{-n}  }{2}\right|^2|B_n|^2G_{A,n}
\\
&\qquad \qquad \quad  
+2\kappa_n(1-\kappa_n)(|A_n|^2+|B_n|^2)(1+2E_n) \Big]\,.
\eal
In this case,
\bal
{\rm var}(\hat r_m)=&
\frac{1}{|A_mB_m^\star+A_{-m}^\star B_{-m}|^2}\cdot\\
&\sum_{n=-N}^N 
\Bigg[
  \left|\frac{\kappa_n+\kappa_{-n}}{2}+{r_m}\right|^2\frac{|A_n|^2}{G_{A,n}}
+ \left|\frac{\kappa_n-\kappa_{-n} }{2}\right|^2|A_n|^2G_{A,n} \\
&\qquad \qquad \quad
+ \left|\frac{\kappa_n+\kappa_{-n} }{2}+{r_m}\right|^2\frac{|B_n|^2}{G_{B,n}}
+ \left|\frac{\kappa_n-\kappa_{-n}  }{2}\right|^2|B_n|^2G_{B,n}
\\
&\qquad \qquad \quad
+ \left|\frac{\kappa_n+\kappa_{-n} }{2}-{r_m}\right|^2\frac{|A_n|^2}{G_{B,n}}
+ \left|\frac{\kappa_n-\kappa_{-n} }{2}\right|^2|A_n|^2G_{B,n} \\
&\qquad \qquad \quad
+ \left|\frac{\kappa_n+\kappa_{-n}  }{2}-{r_m}\right|^2\frac{|B_n|^2}{G_{A,n}}
+ \left|\frac{\kappa_n -\kappa_{-n}  }{2}\right|^2|B_n|^2G_{A,n}
\\
&\qquad \qquad \quad  
+2\kappa_n(1-\kappa_n)(|A_n|^2+|B_n|^2)(1+2E_n) \Bigg].
\label{eq:var_div_cross_supp}
\eal
\end{widetext}
Here we observe that the amplified noises caused by two-mode squeezing mismatching, which undermined the quantum advantage in the intra-comb-line squeezing, now diminish because the cross-comb-line entanglement squeezes both the self-beating noises in the form of $|A_n|^2/G_{A,n}, |B_n|^2/G_{B,n}$ and the cross-beating noises in the form of $|A_n|^2/G_{B,n}, |B_n|^2/G_{A,n}$. Still, the cross-beating noises will completely diminish when $r_m=\kappa$ with uniform absorption spectrum $\kappa_n=\kappa$ for all $n$, similar to the intra-comb-line squeezing case. Nevertheless, here the asymmetry of the transmissivity spectrum $\kappa_n\neq \kappa_{-n}$ leads to antisqueezed quadrature noises $\propto G_{A,n}, G_{B,n}$.

\hw{Now consider uniform squeezing and comb line spectrum $G_{A,n}=G_A,G_{B,n}=G_B$, $A_n=A, B_n=B$ and the single-absorption-line case, i.e. $\kappa_m=\kappa, \kappa_{n\neq m}=1$. Then ${r_m} = (\kappa+1)/2$, the variance Eq.~\eqref{eq:var_div_cross_supp} becomes
\begin{widetext}
\bal
{\rm var}(\hat r_m)&=
\frac{1}{4} \Bigg\{(M-1) \left[\frac{1}{4} (1-\kappa )^2 \left(\frac{1}{A^2 G_A}+\frac{1}{B^2 G_B}\right)+\frac{1}{4} (\kappa +3)^2 \left(\frac{1}{A^2
   G_B}+\frac{1}{B^2 G_A}\right)\right]\\
   &\quad \quad +\frac{1}{4} (1-\kappa )^2 \left(\frac{G_A+G_B}{A^2}+\frac{G_A+G_B}{B^2}\right)+(\kappa +1)^2
   \left(\frac{1}{A^2 G_B}+\frac{1}{B^2 G_A}\right)\\
   &\quad \quad +2
   \kappa  (1-\kappa ) \left(2 E_m+1\right) \left(\frac{1}{A^2}+\frac{1}{B^2}\right)\Bigg\}
   \label{eq:var_div_cross_singleline_supp}
\eal
\end{widetext}
for the division readout at the absorption line.
Take the limit $M\gg G$ and note that the Fisher-type local SNR is defined by Eq.~\eqref{eq:SNR_point} with the signal $ 
\expval{\hat r_m}\simeq {r_m}$ and $|\partial_{\sqrt{\kappa}} \expval{\hat r_m}|^2=\kappa$, we recover Eq.~\eqref{SNR_div_single_line_E} in the maintext.
}

\hw{
We can generalize the above single-absorption-line case to the $\gamma M$-uniform-absorption-line case with uniform transmissivity $\kappa$. We assume $\gamma$ is small and the $\gamma M$ absorption lines locates at the same side of the carrier, then ${r_m} = (\kappa+1)/2$ for all the $\gamma M$ absorption lines. Now the variance Eq.~\eqref{eq:var_div_cross_supp} becomes
\begin{widetext}
\bal
{\rm var}(\hat r_m)&=
\frac{1}{4} \Bigg\{(1-\gamma)M \left[\frac{1}{4} (1-\kappa )^2 \left(\frac{1}{A^2 G_A}+\frac{1}{B^2 G_B}\right)+\frac{1}{4} (\kappa +3)^2 \left(\frac{1}{A^2
   G_B}+\frac{1}{B^2 G_A}\right)\right]\\
   &\quad \quad +\gamma M\left[\frac{1}{4} (1-\kappa )^2 \left(\frac{G_A+G_B}{A^2}+\frac{G_A+G_B}{B^2}\right)+(\kappa +1)^2
   \left(\frac{1}{A^2 G_B}+\frac{1}{B^2 G_A}\right)\right]\\
   &\quad \quad + \gamma M \left[2
   \kappa  (1-\kappa ) \left(2 E_m+1\right) \left(\frac{1}{A^2}+\frac{1}{B^2}\right)\right]\Bigg\}
\label{eq:SNR_div_cross_gammaM_supp}
\eal
\end{widetext}
for the division readouts at all the $\gamma M$ absorption lines.
}

\subsection{Subtraction data processing}

From Eq.~\eqref{eq:I_mf_mean_supp}, it appears that a subtraction between the two photocurrents, in analogy to a balanced homodyne receiver, also yields information about the absorption spectrum. However, different from homodyne, here $\expval{\hat I_B(m\Delta \omega_r)}$ does not carry any information about the sample, so subtraction only invokes an extra unnecessary loss when interfering the information-carrying signal $\hat I_A$ with $\hat I_B$, and it cannot calibrate the unknown comb spectrum like the division data processing. Hence subtraction data processing is not of interest in the combine-then-pass-through case, we skip the derivation for it.

\section{Pass through sample then interfere}
\label{sec:pass_combine}

As shown in Fig.~\ref{fig:schematic}(a) of the main text, the signal comb first passes through the sample and then the combs $\calL[\hat A](t)$ and $\hat B(t)$ interfere at a balanced beamsplitter, leading to output
\bal 
\hat A'(t)&=\frac{\calL[\hat A(t)]+\hat B(t)}{\sqrt{2}}\,,\\
\hat B'(t)&=\frac{\calL[\hat A(t)]-\hat B(t)}{\sqrt{2}}\,.
\eal 
The power on the sample can be obtained as $P_A$ specified in Eq.~\eqref{power_AB}.
At the photon detectors, the time domain photocurrents are
\bal 
\hat I_A(t)
&=\hat A^{\prime\dagger}(t)\hat A'(t)\\
&=\frac{1}{2}(\calL[\hat A(t)]+\hat B(t))^\dagger (\calL[\hat A(t)]+\hat B(t)),
\label{IAt_2}
\eal
\bal 
\hat I_B(t)&=\hat B^{\prime\dagger}(t)\hat B'(t)\\
&=\frac{1}{2}(\calL[\hat A(t)]-\hat B(t))^\dagger (\calL[\hat A(t)]-\hat B(t)).
\label{IBt_2}
\eal
In the data processing, similar to the combine-then-pass case, one performs Fourier transform of the photocurrent and applies a bandpass filter that only keeps the intermediate-frequency components at $\sim\Delta \omega_r$.

To begin with, we solve the modulation on the mean photocurrent. Eqs.~\eqref{IAt_2}\eqref{IBt_2} along with Eqs.~\eqref{eq_combAB_mean}\eqref{eq_combAB_mean_loss} yield
\ba 
&\!\!\!\!\!\!\!\!\!\!\!\!\expval{\hat I_A(t)}&\!=\!\frac{1}{2T}\!\! \sum_{n=-N}^N \!\! e^{in \Delta \omega_rt} \sqrt{\kappa_n}e^{-i\theta_n} A_n^\star B_n\!+\!c.c.\!+\!\cdots
\\
&\!\!\!\!\!\!\!\!\!\!\!\!\expval{\hat I_B(t)}&\!=\!\frac{-1}{2T}\!\! \sum_{n=-N}^N \!\! e^{in \Delta \omega_rt}\sqrt{\kappa_n}e^{-i\theta_n} A_n^\star B_n\!+\!c.c.\!+\!\cdots
\ea 
where `$\cdots$' denote DC term and high-frequency terms to be filtered out. After the bandpass filter, in the frequency domain, we obtain the intermediate-frequency components
\be
\hat I_X(m\Delta \omega_r)\equiv \frac{1}{2\pi }\int e^{im \Delta\omega_r t} \hat I_X(t) dt, 
\ee 
for $X=A,B$.
Their mean values are
\begin{subequations}
\begin{align}
&\expval{\hat I_A(m\Delta \omega_r)}
\nonumber\\
&=\frac{\sqrt{\kappa_m}e^{i\theta_m} A_m B_m^\star+\sqrt{\kappa_{-m}}e^{-i\theta_m} A_{-m}^\star B_{-m}}{2}
\\
&\expval{\hat I_B(m\Delta \omega_r)}
\nonumber\\
&=\frac{-\sqrt{\kappa_m}e^{i\theta_m} A_m B_m^\star-\sqrt{\kappa_{-m}}e^{-i\theta_m} A_{-m}^\star B_{-m}}{2}.
\end{align} 
\label{I_mf_mean_het}
\end{subequations} 

\subsection{Division data processing}
From the calculation of mean values, it is immediately clear that division data processing, which measures the ratio $\hat I_A(m\Delta \omega_r)/\hat I_B(m\Delta \omega_r)$, does not provide any information about the absorption spectrum $\{\kappa_m\}$ to the leading order, because the ratio of the mean values is a constant $-1$ independent on $\{\kappa_m\}$. In this sense, we only analyze the subtraction data processing. 

\subsection{Subtraction data processing}
Subtraction data processing measures the differential photocurrent spectrum
\be 
\hat{d}_m=\hat I_A(m\Delta \omega_r)-\hat I_B(m\Delta \omega_r).
\label{eq:d_m_het_supp}
\ee 
Its mean value is
\be 
\expval{\hat{d}_m}=\sqrt{\kappa_m}e^{i\theta_m} A_m B_m^\star+\sqrt{\kappa_{-m}}e^{-i\theta_m} A_{-m}^\star B_{-m}.
\label{d_m_het_app}
\ee 
The variance is
\begin{align}
{\rm var}(\hat{d}_m)={\rm var}[\Delta\hat I_A(m\Delta \omega_r)-\Delta\hat I_B(m\Delta \omega_r)]\,.
\end{align}
From Eq.~\eqref{eq:SNR_point} of the main text, the SNR for the estimation of $\sqrt{\kappa_m}$ is given by $|A_mB_m^\star|^2/{\rm var}(\hat{d}_m)$. On the other hand, the global SNR, defined in Eq.~\eqref{eq:SNR_global} of the main text, is $|\expval{\hat{d}_m}(\kappa)-\expval{\hat{d}_m}(\kappa=1)|^2/{\rm var}(\hat{d}_m)$.

Now we solve the variance explicitly. We immediately note that the difference has a simple form
\be 
\hat{I}_A(t)-\hat{I}_B(t)=\calL[\hat A(t)]^\dagger \hat B(t)+\hat B^\dagger(t)\calL[\hat A(t)].
\ee 
Therefore, we evaluate the difference directly
\begin{align}
&\Delta\hat I_A(t)-\Delta\hat I_B(t)=
\nonumber
\\
&\expval{\calL[\hat A(t)]}^\dagger\Delta \hat B(t)+ \expval{\hat B(t)}^\dagger \Delta \calL[\hat A(t)]+h.c.+\cdots,
\nonumber\\
&=\frac{1}{T}\sum_{n=-N}^N\sum_{m=-2N}^{2N} \left[e^{i (n-m)\Delta\omega_r t} \sqrt{\kappa_n } A_n^\star e^{-i\theta_n}   \hat \calB_{n,m}+\right.
\nonumber\\
& \qquad \quad  \left.B_n^\star  e^{-i (m\Delta\omega_r) t}\left(\sqrt{\kappa_n} e^{i\theta_n} \hat \calA_{n,m}+\sqrt{1-\kappa_n}\hat\calV_{n,m} \right)\right]
\nonumber\\
&\quad +h.c.+\cdots
\label{Delta_IAt_2}
\end{align}
where we have used Eqs.~\eqref{eq_combAB}, ~\eqref{eq_combAB_mean} and~\eqref{eq_combAB_mean_loss}, `$\cdots$' denote DC term and high-frequency terms to be filtered out.
Finally, we have
\begin{align} 
&\Delta \hat{d}_m=
\nonumber
\\
&\sum_{n=-N}^N \sqrt{\kappa_n } A_n^\star e^{-i\theta_n}   \hat \calB_{n,n+m}
\nonumber\\
&\quad +B_n^\star  \left(\sqrt{\kappa_n} e^{i\theta_n} \hat \calA_{n,m}+\sqrt{1-\kappa_n}\hat\calV_{n,m} \right)
\nonumber
\\
&\quad +\sqrt{\kappa_n } A_n e^{i\theta_n}   \hat \calB_{n,n-m}^\dagger
\nonumber\\
&\quad +B_{n}\left(\sqrt{\kappa_n} e^{-i\theta_n} \hat \calA_{n,-m}^\dagger+\sqrt{1-\kappa_n}\hat\calV_{n,-m}^\dagger \right)
\label{eq:d_noise_supp}
\end{align}

\subsubsection{Intra-comb-line squeezing}
To reduce the noise ${\rm var}(\hat d_m)$, first we consider the intra-line sideband two-mode squeezings between $\{(\hat \calB_{n,n+m}, \hat \calB_{n,n-m})\}$ for comb $B$ and $(\hat\calA_{n,m}, \hat\calA_{n,-m})$ for comb $A$, which was first proposed in Ref.~\cite{shi2023entanglement}. 

Assuming real comb amplitude $A_n=A_n^*$, $B_n=B_n^*$, we have
\begin{align} 
&{\rm var} (\hat{d}_m)=
\nonumber
\\
&\sum_{n=-N}^N \kappa_n A_n^2 {\rm var } \left(e^{-i\theta_n}   \hat \calB_{n,n+m}
+ e^{i\theta_n}   \hat \calB_{n,n-m}^\dagger\right)
\nonumber
\\
&\quad + \kappa_n B_n^2 {\rm var}
\left(e^{i\theta_n} \hat \calA_{n,m}+
 e^{-i\theta_n} \hat \calA_{n,-m}^\dagger\right)\\
&\quad + (1-\kappa_n) B_{n}^2 {\rm var}\left(\hat\calV_{n,m}+\hat\calV_{n,-m}^\dagger \right)
\end{align}
Consider input modes in two-mode squeezed state with ${\rm var} (\hat \calB_{n,n+m}+ \hat \calB_{n,n-m}^\dagger)=1/{G_{B,n}}$, ${\rm var}(\hat\calA_{n,m}+ \hat\calA_{n,-m}^\dagger)=1/{G_{A,n}}$, and environment mode in thermal state with ${\rm var}(\hat\calV_{n,m}+ \hat\calV_{n,-m}^\dagger)=1+2E_n$, assuming phase corrected $\theta_n=0$, we have
\begin{align}
{\rm var}(\hat{d}_m|\{\theta_n=0\}_n)&=\sum_{n=-N}^N 
 \kappa_n \left(\frac{A_n^2}{G_{B,n}}  +  \frac{B_n^2}{G_{A,n}} \right) 
 \nonumber
 \\
 &+ (1-\kappa_n) B_{n}^2  (1+2E_n).
\end{align}
We use this formula to produce the numerical evaluations of SNR in the maintext.

If $\theta_n\neq 0$, the antisqueezed (amplified) noise is involved:
\bal 
\!&{\rm var}
\left(e^{i\theta_n} \hat \calA_{n,m}+
 e^{-i\theta_n} \hat \calA_{n,-m}^\dagger\right)=\\
 &\frac{1}{ 2G_{A,n}}\left[-\left(G_{A,n}^2-1\right) \cos \left(2 \theta_n \right)+(G_{A,n}^2+1)\right]\,,
\eal
~
\bal 
\!&{\rm var } \left(e^{-i\theta_n}   \hat \calB_{n,n+m}
+ e^{i\theta_n}   \hat \calB_{n,n-m}^\dagger\right)=\\
 &\frac{1}{ 2G_{B,n}}\left[-\left(G_{B,n}^2-1\right) \cos \left(2 \theta_n \right)+(G_{B,n}^2+1)\right]\,.
\eal
In this case
\begin{align} 
&{\rm var} (\hat{d}_m)=
\nonumber
\\
&\sum_{n=-N}^N \kappa_n A_n^2 \frac{1}{ 2G_{B,n}}\left[-\left(G_{B,n}^2-1\right) \cos \left(2 \theta_n \right)+(G_{B,n}^2+1)\right]
\nonumber
\\
&\quad + \kappa_n B_n^2 
\frac{1}{ 2G_{A,n}}\left[-\left(G_{A,n}^2-1\right) \cos \left(2 \theta_n \right)+(G_{A,n}^2+1)\right]
\nonumber
\\
&\quad + (1-\kappa_n) B_{n}^2 (1+2E_n).
\label{eq:var_het_intra_supp}
\end{align}

\newpage

\hw{Now consider uniform squeezing and comb line spectrum $G_{A,n}=G_A,G_{B,n}=G_B$, $A_n=A, B_n=B$ and the single-absorption-line case, i.e. $\kappa_m=\kappa, \kappa_{n\neq m}=1$. Then 
the variance Eq.~\eqref{eq:var_het_intra_supp} becomes
\begin{widetext}
\bal
{\rm var}(\hat d_m)&=
(M-1) \left(\frac{A^2}{G_B}+\frac{B^2}{G_A}\right)+\frac{A^2 \kappa }{G_B}+\frac{B^2 \kappa }{G_A}+B^2 (1-\kappa ) \left(2 E_m+1\right)
   \label{eq:var_het_intra_singleline_supp}
\eal
\end{widetext}
for the heterodyne readout at the absorption line.
Take the limit $M\gg G$ and note that the Fisher-type local SNR is defined by Eq.~\eqref{eq:SNR_point} with the signal $\expval{\hat d_m} =(\sqrt{\kappa_m} +1)AB$ and $|\partial_{\sqrt{\kappa}} \expval{\hat d_m}|^2=|AB|^2$, we recover Eq.~\eqref{SNR_het_singleline} in the maintext.
}

\hw{
We can generalize the above single-absorption-line case to the $\gamma M$-uniform-absorption-line case with uniform transmissivity $\kappa$.
Now the variance Eq.~\eqref{eq:var_het_intra_supp} becomes
\begin{widetext}
\bal
{\rm var}(\hat d_m)&=
(1-\gamma) M \left(\frac{A^2}{G_B}+\frac{B^2}{G_A}\right)+\gamma M \left[\frac{A^2 \kappa }{G_B}+\frac{B^2 \kappa }{G_A}+B^2 (1-\kappa ) \left(2 E_m+1\right)\right]
\label{eq:SNR_het_intra_gammaM_supp}
\eal
\end{widetext}
for the heterodyne readouts at all the $\gamma M$ absorption lines.
}

\subsubsection{Cross-comb-line entanglement}
Alternatively, we can consider cross comb line entanglement, generated by a single pump line at the carrier frequency, which has been adopted in Ref.~\cite{hariri2025entangled}.
Formally, it implements two-mode squeezing between the pairs of $(\hat \calB_{n,n+m},\hat \calB_{-n,-n-m})$ and $(\hat \calA_{n,m}, \hat \calA_{-n,-m})$. Below we rewrite the readout noise $\Delta \hat d_m$ in such frequency pairs around the carrier frequency. 
\begin{widetext}
\begin{align} 
\Delta \hat d_m&= \left(\sqrt{\kappa_0 } A_0^\star e^{-i\theta_0}   \hat \calB_{0,0+m}\right)
+ \left(B_0^\star \sqrt{\kappa_0} e^{i\theta_0} \hat \calA_{0,m}  \right) 
\nonumber
\\
&\qquad 
+\left(\sqrt{\kappa_0 } A_0 e^{i\theta_0}   \hat \calB_{0,0-m}^\dagger\right)
+ \left(B_{0} \sqrt{\kappa_0} e^{-i\theta_0} \hat \calA_{0,-m}^\dagger\right)
\nonumber
\\
&\qquad 
+B_{0} \sqrt{1-\kappa_0}\hat\calV_{0,-m}^\dagger 
+B_0^\star \sqrt{1-\kappa_0}\hat\calV_{0,m}
\nonumber
\\
&\quad \quad +\sum_{n=1}^N 
\Bigg\{\!\!
\left(\!\sqrt{\kappa_n } A_n^\star e^{-i\theta_n}   \hat \calB_{n,n+m} \!+\! \sqrt{\kappa_{-n} } A_{-n} e^{i\theta_{-n}}   \hat \calB_{-n,-n-m}^\dagger\right)
\!\!+\!\! \left(\! B_n^\star \sqrt{\kappa_n} e^{i\theta_n} \hat \calA_{n,m} \!+\! B_{-n}\sqrt{\kappa_{-n}} e^{-i\theta_{-n}} \hat \calA_{-n,-m}^\dagger \right) 
\nonumber
\\
&\quad \qquad 
+\left(\sqrt{\kappa_n } A_n e^{i\theta_n}   \hat \calB_{n,n-m}^\dagger+ \sqrt{\kappa_{-n} } A_{-n}^\star e^{-i\theta_{-n}}   \hat \calB_{-n,-n+m}\right)
\!+\! \left(B_{n} \sqrt{\kappa_n} e^{-i\theta_n} \hat \calA_{n,-m}^\dagger + B_{-n}^\star \sqrt{\kappa_{-n}} e^{i\theta_{-n}} \hat \calA_{-n,m}\right)
\nonumber
\\
&\quad \qquad 
+B_{-n}^\star \sqrt{1-\kappa_{-n}}\hat\calV_{-n,m}
+B_{n} \sqrt{1-\kappa_n}\hat\calV_{n,-m}^\dagger 
+B_{-n}  \sqrt{1-\kappa_{-n}}\hat\calV_{-n,-m}^\dagger 
+B_n^\star \sqrt{1-\kappa_n}\hat\calV_{n,m}\Bigg\}\,.
\end{align}
\end{widetext}

Now consider the noises of EPR quadratures defined in Eq.~\eqref{eq:EPR_quadrature_noise}. Then following the same procedure of $\hat\Sigma_Q, \hat\Sigma_P$ noise decomposition in obtaining Eq.~\eqref{eq:CrossEntangle_var_QP_app}, we derive the readout noise 
\clearpage
\ba
&~&{\rm var}(\hat d_m)  ={\rm var}(\Re\hat d_m)+{\rm var}(\Im\hat d_m)
\nonumber\\
&&= \frac{\kappa_0|A_0|^2}{G_{B,0}(\theta_0)} + \frac{\kappa_0|B_0|^2}{G_{A,0}(\theta_0)} +\sum_{n=-N}^N  (1-\kappa_{n})|B_n|^2 (1+2E_n)
\nonumber\\
&&\quad
+\sum_{n=1}^N  \frac{1}{2}\Big[  |\sqrt{\kappa_n}A_n^* e^{-i\theta_n} -\sqrt{\kappa_{-n}} A_{-n}e^{i\theta_{-n}} |^2G_{B,n}
\nonumber\\
&&\qquad \qquad+   |\sqrt{\kappa_n}A_n^*e^{-i\theta_n} +\sqrt{\kappa_{-n}} A_{-n}e^{i\theta_{-n}} |^2/G_{B,n}
\Big]
\nonumber\\
&&\quad
+\sum_{n=1}^N  \frac{1}{2}\Big[  |\sqrt{\kappa_n}B_n^*e^{i\theta_n} -\sqrt{\kappa_{-n}} B_{-n}e^{-i\theta_{-n}} |^2G_{A,n}
\nonumber\\
&&\qquad \qquad
+   |\sqrt{\kappa_n}B_n^* e^{i\theta_n}+\sqrt{\kappa_{-n}} B_{-n} e^{-i\theta_{-n}}|^2/G_{A,n}
\Big]
\,,
\nonumber\\
\label{eq:var_het_cross_supp}
\ea
where 
\bal 
\!G_{A,0}(\theta_0)\equiv&{\rm var } \left(e^{i\theta_0}   \hat \calA_{0,m}
+ e^{-i\theta_{0}}   \hat \calA_{0,-m}^\dagger\right)\\
 =&\frac{1}{ 2G_{A,0}}\left[-\left(G_{A,0}^2-1\right) \cos \left(2 \theta_0 \right)+(G_{A,0}^2+1)\right]\,,
\eal
\bal 
\!G_{B,0}(\theta_0)\equiv&{\rm var } \left(e^{-i\theta_0}   \hat \calB_{0,m}
+ e^{i\theta_{0}}   \hat \calB_{0,-m}^\dagger\right)\\
 =&\frac{1}{ 2G_{B,0}}\left[-\left(G_{B,0}^2-1\right) \cos \left(2 \theta_0 \right)+(G_{B,0}^2+1)\right]\,.
\eal
We use this formula under $\theta_0=0$ to produce the numerical evaluations of SNR in the maintext.

An experiment relevant scenario is the strong LO limit $|B_n|\gg |A_n|$ such that only the noises from $\calA$ dominates. For simplicity, let us assume phase known and compensated, $\theta_n=0$ for all $-N\le n\le N$, $B_n$ are real. In this case, the readout noise reduces to
\ba
&~&{\rm var}(\hat d_m|\{\theta_n=0,B_n\gg A_n\}_n)  
\nonumber\\
&&= \frac{\kappa_0B_0^2}{G_{A,0}} +\sum_{n=-N}^N  (1-\kappa_{n})B_n^2 
\nonumber\\
&&\quad+\sum_{n=1}^N  \frac{1}{2}\Big[  (\sqrt{\kappa_n}B_n -\sqrt{\kappa_{-n}} B_{-n} )^2G_{A,n}
\nonumber\\
&&\qquad \qquad+   (\sqrt{\kappa_n}B_n +\sqrt{\kappa_{-n}} B_{-n} )^2/G_{A,n}
\Big]\,.
\ea

\hw{Now consider uniform squeezing and comb line spectrum $G_{A,n}=G_A,G_{B,n}=G_B$, $A_n=A, B_n=B$ and the single-absorption-line case, i.e. $\kappa_m=\kappa, \kappa_{n\neq m}=1$. Then 
the variance Eq.~\eqref{eq:var_het_cross_supp} becomes
\begin{widetext}
\bal
{\rm var}(\hat d_m)&=
\frac{G_A \left| \sqrt{\kappa }-1\right| ^2+\frac{\left| \sqrt{\kappa }+1\right| ^2}{G_A}}{2 }B^2+\frac{M-2}{ G_A}B^2
+\frac{\left| \sqrt{\kappa }-1\right| ^2 G_B+\frac{\left| \sqrt{\kappa }+1\right| ^2}{G_B}}{2 }A^2+\frac{M-2}{G_B}A^2
   +(1-\kappa ) \left(2 E_m+1\right)B^2
   \label{eq:var_het_cross_singleline_supp}
\eal
\end{widetext}
for the heterodyne readout at the absorption line.
Take the limit $M\gg G$ and note that the Fisher-type local SNR is defined by Eq.~\eqref{eq:SNR_point} with the signal $\expval{\hat d_m} =(\sqrt{\kappa_m} +1)AB$ and $|\partial_{\sqrt{\kappa}} \expval{\hat d_m}|^2=|AB|^2$, we recover Eq.~\eqref{SNR_het_singleline} in the maintext.
}

\hw{
We can generalize the above single-absorption-line case to the $\gamma M$-uniform-absorption-line case with uniform transmissivity $\kappa$.
We assume $\gamma$ is small and the $\gamma M$ absorption lines locates at the same side of the carrier, then each of the $\gamma M$ absorption lines affects only one noise mode of the noise pair centered around the carrier. 
Now the variance Eq.~\eqref{eq:var_het_cross_supp} becomes
\begin{widetext}
\bal
{\rm var}(\hat d_m)&=
\gamma  M\left[\frac{G_A \left| \sqrt{\kappa }-1\right| ^2+\frac{\left| \sqrt{\kappa }+1\right| ^2}{G_A}}{2 }B^2+\frac{\left| \sqrt{\kappa }-1\right| ^2 G_B+\frac{\left| \sqrt{\kappa
   }+1\right| ^2}{G_B}}{2 }A^2
   +(1-\kappa ) \left(2 E_m+1\right)B^2\right]\\
   &\quad +\frac{2 \left(\frac{M-1}{2}-\gamma  M\right)+1}{G_A}B^2+\frac{2 \left(\frac{M-1}{2}-\gamma  M\right)+1}{G_B}A^2
\label{eq:SNR_het_cross_gammaM_supp}
\eal
\end{widetext}
for the heterodyne readouts at all the $\gamma M$ absorption lines.
}

\hw{
\section{Decoding the aliasing between the positive- and negative-indexed comb lines}
\label{sec:aliase_supp}
\subsection{Two-shot measurement without prior knowledge}
To decode the absorption on the positive- and negative-index comb lines, i.e., $\sqrt{\kappa_m} e^{i\theta_m}$ and $\sqrt{\kappa_{-m}}e^{-i\theta_{-m}}$ concurrently, one adopts a two-shot measurement protocol. 
}

\hw{
First consider division receiver. The first shot produces division ratio with mean value Eq.~\eqref{eq:rm_app} as 
\be 
r_m^{(1)}= \frac{A_{m} B_{m}^\star  \kappa_m+ A_{-m}^\star B_{-m} \kappa_{-m} }{A_mB_m^\star+A_{-m}^\star B_{-m}}
\ee
In the second shot, a phase shift of $\pi$ is applied to the positive index comb lines of comb $A$ only, yielding $A_m\to -A_m $ for all $1\le m\le N$ prior to combining it with the entangled comb. The second division measurement produces mean value 
\be 
r_m^{(2)}= \frac{-A_{m} B_{m}^\star \kappa_m + A_{-m}^\star B_{-m} \kappa_{-m} }{-A_mB_m^\star+A_{-m}^\star B_{-m}}
\ee
Subsequently, two linear combinations of $\kappa_{m}, \kappa_{-m}$ are obtained, which allow to solve for both spectral components.
}

\hw{
Similarly, for heterodyne detection, the first shot produces the differential photocurrent with mean value Eq.~\eqref{d_m_het_app} as
\begin{align}
\expval{\hat d^{(1)}_m}
=\sqrt{\kappa_m} e^{i\theta_m}  A_m B_m^* + \sqrt{\kappa_{-m}} e^{-i\theta_{-m}} A_{-m}^* B_{-m}
\,,
\end{align} 
In the second shot, a positive-sideband phase shift of $\pi$ is applied yielding $A_m\to -A_m $ for all $1\le m\le N$ prior to combining it with the entangled comb. The heterodyne measurement produces photocurrent
\begin{align}
\expval{\hat d^{(2)}_m}
= - \sqrt{\kappa_m} e^{i\theta_m} A_m B_m^*+ \sqrt{\kappa_{-m}} e^{-i\theta_{-m}} A_{-m}^*  B_{-m}
\,.
\end{align}
Subsequently, two linear combinations of $\sqrt{\kappa_{m}}e^{i\theta_m}, \sqrt{\kappa_{-m}}e^{-i\theta_{-m}}$ are obtained, which allow to solve for both spectral components.
}

\hw{
\subsection{Single-sided absorption}
If the absorption is known to be on only positive- or negative-index comb lines, a single measurement suffices to infer the absorption spectrum. Without loss of generality, we assume the negative-index comb lines are lossless, i.e., $\kappa_{-m}=1$ for $1\le m\le N$.
}

\hw{
For division receiver, the mean value of division ratio reduces to
\be 
r_m= \frac{A_{m} B_{m}^\star  \kappa_m+ A_{-m}^\star B_{-m} }{A_mB_m^\star+A_{-m}^\star B_{-m}}\,.
\ee
If $A_m=A_{-m}=A$, $B_m=B_{-m}=B$, then
\be 
r_m=(1+\kappa_{m})/2\,.
\ee
}

\hw{
For heterodyne detection, the mean value of the differential photocurrent reduces to
\begin{align}
&\expval{\hat d_m}
= \sqrt{\kappa_{m}} e^{i\theta_m} A_m B_m^* + A_{-m}^* B_{-m}
\,.
\end{align} 
If $A_m=A_{-m}=A$, $B_m=B_{-m}=B, \theta_m=0$, then
\be 
\expval{\hat d_m}=(1+\sqrt{\kappa_{m}})AB\,.
\ee
\subsection{One-shot measurement with known phase}
When the absorption phase spectrum $\{\theta_m\}_{m=-N}^N$ is known or small, one lets  $\{\theta_m = 0\}_{m=-N}^N$ without loss of generality. Further assuming $A_m$'s are real for $1<m\leq N$ and pure imaginary for $-N \leq m < -1$, and $B_m$'s are real, which can be attained by the waveshaper, the positive- and negative-index comb lines can be independently decoded in the real and imaginary components of the readout mean value:
for division receiver
\be 
r_m= \frac{A_{m} B_{m}  \kappa_m+ i |A_{-m}| B_{-m} \kappa_{-m} }{A_mB_m+i|A_{-m}| B_{-m}}
\ee
and for heterodyne detection
\be
\expval{\hat d_m}
=\sqrt{\kappa_m}A_m B_m  +i\sqrt{\kappa_{-m}}|A_{-m}|B_{-m}\,.
\ee
Remarkably, the prior knowledge about the common phase is not necessary for division receiver, since it is only sensitive to differential phases between $A$ and $B$.
}

\section{Influence of phase noise} 
\label{sec:phase_noise_supp}
\QZ{
We consider the same single-comb line absorption detection problem, however, with random phase noise across all modes $\theta_n\in[-\delta,\delta]$ being unknown. We will assume the phase noise being small, $\delta\ll1$. The goal is to estimate the single absorption line $\kappa<1$ in the presence of unknown phase noise.
}

\QZ{
The division receiver has the same SNR as in Eq.~\eqref{SNR_div_single_line} of the main text. As we mentioned, division receiver cannot measure phase, and at the same time is robust to phase noise. 
}

\QZ{
For heterodyne based scheme, we will focus on the intra-comb-line squeezing for simplicity. In heterodyne detection, one can estimate the sample induced absorption and phase simultaneously in Eq.~\eqref{I_mf_mean_het}. If there is a fixed phase mismatching $\delta$, it will lead to an increase in the variance, assuming uniform mismatching $\delta$ over all frequencies,
\begin{align} 
&{\rm SNR}_{\rm het}^{-2}\simeq
\frac{M}{A^2B^2}\Big\{ \frac{A^2}{ 2G_{B}}\left[-\left(G_{B}^2-1\right) \cos \left(2 \delta \right)+(G_{B}^2+1)\right]
\nonumber
\\
&\quad +  
\frac{B^2}{ 2G_{A}}\left[-\left(G_{A}^2-1\right) \cos \left(2 \delta \right)+(G_{A}^2+1)\right]\Big\}.
\label{eq:phase_noise_supp}
\end{align}
}
\hw{Such excess noise can be corrected by adaptive estimation of $\delta$ to compensate it to zero. On the other hand, if there is a weak phase noise with phase variance $\sim\delta$, one can use Eq.~\eqref{eq:phase_noise_supp} to approximate the resulting SNR decrease given $\delta\ll 1$.
}



\end{document}